\documentclass[apjl]{emulateapj}
\usepackage{epsfig}

\begin{document}

\title{
Reducing the Probability of Capture into Resonance
}

\author{Alice C. Quillen}
\affil{Department of Physics and Astronomy, University of Rochester, Rochester, NY 14627}
\affil{Visitor, Research School of Astronomy and Astrophysics, 
Australian National University, 
Mount Stromlo Observatory, Cotter Road, Weston Creek, ACT 2611, Australia
}
\email{aquillen@pas.rochester.edu}

\begin{abstract}
A migrating planet can capture planetesimals
into mean motion resonances.   
However, resonant trapping can be prevented when the drift or migration rate
is sufficiently high.
Using a simple Hamiltonian system for first and second order resonances,
we explore how the capture probability depends
on the order of the resonance, drift rate
and initial particle eccentricity.  
We present scaling factors as a function of the planet mass
and resonance strength
to estimate the planetary migration rate above which 
the capture probability drops to less than 1/2. 
Applying our framework to multiple extra
solar planetary systems that have
two planets locked in resonance, we estimate lower limits for the 
outer planet's migration rate allowing resonance capture
of the inner planet.

Mean motion resonances are comprised of multiple resonant subterms.
We find that the corotation subterm 
can reduce the probability of capture when the planet
eccentricity is above a critical value.  
We present 
factors that can be used to estimate this critical planet eccentricity.
Applying our framework to the migration of Neptune,
we find that Neptune's eccentricity
is near the critical value that would make its 2:1 resonance
fail to capture twotinos.
The capture probability is affected by the separation
between resonant subterms and so is also a function of the precession rates
of the longitudes of periapse of both planet and particle near resonance.


\end{abstract}


\section{Introduction}

Resonances can capture particles in 
slowly varying dynamical systems. 
For example, a planet migrating outward
can trap planetesimals in resonances exterior to it; as Neptune 
trapped the Plutinos in the Kuiper belt 
(e.g., \citealt{fernandez84,malhotra95,hahn99,ida00,chiang02,zhou02,levison03,wyatt03}).
A planet migrating inward can trap planetesimals or planets
in resonances interior to it (e.g., \citealt{quillenholman00,kley04}).
Dust spiraling inward under dissipational forces can become trapped
in exterior mean motion resonances with a planet
(e.g., \citealt{sicardy93,dermott94,marzari94,liou97,liou99,ozernoy00,wilner02,moromartin05,deller05}).

An elegant and predictive theory of resonant capture has been developed for 
adiabatically varying non-chaotic integrable
resonant systems \citep{yoder79,henrard82,henrard83,malhotra90},
and for the same systems but varying 
with weak nonconservative forces \citep{gomes97}.
This theory was first applied to systems of tidally locked satellites 
\citep{borderies84,peale86,dermott88,malhotra90}.
However this theory does not apply to systems that are near
or in the non-adiabatic regime, or are chaotic.
Numerical explorations of drifting and migrating systems
have revealed differences between measured
capture probabilities and those
predicted by the adiabatic theory.
For dust drifting under dissipational forces, small particles
can be drifting sufficiently fast that they are in 
the non-adiabatic regime.  In this case, the capture probability
is reduced (e.g., \citealt{gomes95,liou99}).
Numerical simulations of Neptune's migration show
that if Neptune migrates rapidly, the capture probability
of resonances is reduced \citep{ida00,friedland01,chiang02}.

In dynamical systems chaotic motion can arise from the overlap
of resonances (e.g., \citealt{wisdom80}).
Mean motion resonances associated with perturbations by a planet
are made up of more than one resonant perturbative term.
\citet{holman96} showed how to predict the Lyapunov times of asteroids
from the overlap of these resonant terms.
The complex behavior 
of chaotic resonances could influence the way they capture particles
when they are varying 
(e.g., \citealt{dermott88,tittemore90,sicardy93,marzari94,quillen_cap}).
For example, \citet{tittemore90} found that the drifting chaotic 
resonances of the Uranian satellites exhibited 
different behavior than non-chaotic resonances.

In this paper we strive to develop a general 
framework that  will allow us to better predict the capture
probability of resonances. We would like to
know when they are likely to capture and how
the capture probability depends on the drift or
migration rate, resonance order and strength, initial 
particle orbital elements, and size and separation of subresonances.
Some of these parameters depend on the planetary properties
and others depend on the particle properties.
We would like a flexible framework that could be used to place
constraints on migrating planetary extrasolar systems and the properties
of the planets and planetesimals residing within them.
Previous works have shown how to
predict the capture probability as a function
of initial particle eccentricity in the adiabatic limit  
(e.g., \citealt{henrard82,borderies84,malhotra90,M+D}).
Here we strive to understand two regimes that have not been 
well explored for mean motion resonances.
We probe the nonadiabatic regime 
in which the drift rate is so fast that the resonance can fail
to capture.  We also explore resonances 
comprised of multiple resonant subterms. 

Our approach is to understand the simplest Hamiltonian model
that can illustrate the dynamics of resonance capture.
In section 2 we formulate the problem in terms of an idealized
Hamiltonian with multiple resonant terms.
By rescaling, this simple model 
allows different resonant systems to be treated in a similar way.
In section 3 we consider the system with one resonant
term and measure the capture probability
as a function of drift rate and initial particle momentum. 
We extend previous analytical work by \citet{friedland01} for
the first order resonances and find general expressions
that account for trends exhibited by previous
simulations (e.g., \citealt{wyatt03}).
In section 4 we measure capture probabilities for 
first and second order resonances containing
multiple resonant terms.
In section 5 we illustrate how our framework can 
be applied to planetary systems.
A summary and conclusion follows.
The appendix lists coefficients for common resonances so that
drift rates can be predicted and compared to numerical studies.

\section{Hamiltonian Formulation}

We employ the Poincar\'e coordinates
\begin{displaymath} 
\lambda = M + \varpi, \qquad \gamma = -\varpi 
\end{displaymath}
and their associated momenta
\begin{displaymath}
L = \sqrt{GM_* a}, \qquad \Gamma  = \sqrt{GM_* a}(1-\sqrt{1-e^2})
\end{displaymath}
where $M_*$ is the mass of the star,
$\lambda$ is the mean longitude, $M$ is the mean anomaly, $\varpi$
is the longitude of pericenter, $a$ is the semimajor axis, and $e$ is
the eccentricity. These variables are those
describing the orbit of a particle or planetesimal in a plane.  
The Hamiltonian for the Keplerian system in these coordinates restricted
to a plane is
\begin{displaymath}
H(L,\lambda;\Gamma, \gamma) = -{GM_* \over 2 L^2} - R 
\end{displaymath}
where $R$ is the disturbing function that depends on the
coordinates of the particle and on the coordinates
of the planet.  The planet's semimajor axis and mass are denoted $a_p$
and $m_p$, respectively. The planet's other coordinates
are subscripted in the same way.
The mean motion of the particle $n = \dot \lambda$ where 
$\dot \lambda$ is the derivative with respect to time of $\lambda$.

Hereafter we adopt a unit convention with distances 
in units of the planet's semi-major axis, $a_p$, at a time $t_0$.
Time is put in units of $\sqrt{a_p^3/GM_*}$.
We define $\mu$ to be the mass ratio $\mu \equiv m_p /M_*$.
At low eccentricity, $\Gamma/L \approx e^2/2$, relating the momentum $\Gamma$
to the particle eccentricity.
We often give the particle semimajor axis in terms of
the variable $\alpha \equiv {a_p \over a}$ if $a>a_p$ 
(external to the planet)
and $\alpha \equiv {a \over a_p}$ for $a<a_p$ (internal to the planet).

The unperturbed Hamiltonian or that lacking the disturbing function
\begin{displaymath}
H_0(L,\lambda; \Gamma, \gamma)  = -{1\over 2 L^2} 
\end{displaymath}
We consider the the $j:j-k$ exterior mean motion resonance (planet
is an interior perturber).
We perform a canonical transformation using the mixed variable 
generating function
\begin{displaymath}
F_2 = I (j \lambda - (j-k)\lambda_p) 
\end{displaymath}
leading to new variables
\begin{displaymath}
I =  j L, \qquad  \psi =  j\lambda - (j-k) \lambda_p 
\end{displaymath}
and new Hamiltonian
\begin{displaymath}
H_0'(I , \psi; \Gamma, \gamma) 
  =  {- 1 \over 2 j^2  I^2} -  {(j -k) } I n_p.
\end{displaymath}

We now expand around the resonance.
Let 
\begin{equation}
\Lambda \equiv  I - I_0 
\label{eqnI0}
\end{equation}
and
\begin{displaymath}
{ 1 \over j^2 I_0^3} = {(j-k)} n_p(t_0).
\end{displaymath}
Since we have adopted units $n_p(t_0) = 1$, we find 
$I_0 = \alpha^{-1/2}/j$ where 
$\alpha = {a_p \over a}= 
\left({j-k\over j}\right)^{2/3}$ on resonance, as expected.
Our Hamiltonian now reads
\begin{displaymath}
K_0(\Lambda, \psi; \Gamma, \gamma) =  
   {\rm constant \ } -  (j -k)(n_p-1)\Lambda
 - {3 \Lambda^2 \over 2 j^2 I_0^4}. \nonumber
\end{displaymath}
We can write the unperturbed Hamiltonian as
\begin{displaymath}
K_0(\Lambda, \psi; \Gamma, \gamma) 
  = {a} \Lambda^2 + b \Lambda  + {\rm constant}
\end{displaymath}
with coefficients
\begin{eqnarray}
a &=& -{3 \over 2} j^2 \alpha^{2}      \nonumber \\
b &=& -(j-k) (n_p - 1).
\label{eqnab}
\end{eqnarray}

We now recover the disturbing function that is  
is traditionally expanded as a cosine series of angles in orders
of planet and particle eccentricity.
We keep the terms inducing precession of the longitude
of periapse and low order terms (in eccentricity) containing $\psi$.
The full Hamiltonian
\begin{eqnarray}
K(\Lambda, \psi; \Gamma,\gamma) 
   &  = & 
   a \Lambda^2 + b \Lambda + c \Gamma 
  \qquad  
\label{hamgen}
\\
& &  + \sum_{p=0}^k \delta_{k,p}    \Gamma^{(k-p)/2} 
      \cos{( \psi - (k-p) \varpi - p \varpi_p)}   
\nonumber
\end{eqnarray}
with coefficient
\begin{equation}
c        = -{\mu 2 f_2 \alpha^{1/2}}.
\label{eqnc}
\end{equation}
We have used the approximation $e^2 \sim 2 \Gamma/L 
\sim 2 \Gamma \alpha^{1/2}$.
Here the perturbation strengths, $\delta_{k,p}$, are
functions of $\alpha$, $j$ and Laplace coefficients (see \citealt{M+D}).
The $c$ term describes secular precession of the longitude of periapse
and depends on the function $f_2$ given in 
the appendix by \citet{M+D} and is evaluated at $\alpha$ with index 
$j=0$.
As have previous studies, 
we have neglected the dependence of $\alpha$ on time as the planet
migrates,
and cosine terms from the disturbing function expansion
which are expected to average to zero near resonance 
(e.g., \citealt{borderies84,peale86,holman96,M+D}). 
The perturbation strengths depend on the planet
mass and eccentricity as $\delta_{k,p} \propto \mu e_p^p$.
More detailed expressions are listed in the appendix.

The above canonical transformations are similar to those
of \cite{holman96} except we have focused on resonances
exterior to a planet rather than those interior to it.
We have also explicitly kept the $b \Lambda$ term.
A system with a migrating planet would be described
by a time dependent $b$ coefficient.
This allows us to explore the dynamical behavior
as particles pass through resonances.

\citet{holman96} showed that
the above Hamiltonian is similar to a periodically forced 
pendulum and that the overlap of the different resonant terms
of Equation (\ref{hamgen})
can induce largescale chaotic behavior.
However most previous explorations of resonant capture
have only considered one dominant resonant term.
When the migration rate is slow, the adiabatic theory
developed by \citep{yoder79,henrard82,borderies84,malhotra90,M+D} 
applies.   In the next section we explore this simpler
situation, but allow the migration rate to be fast or non-adiabatic.

\section{Probability of Capture in a Single Resonance}

In this section we explore the simpler
Hamiltonian containing only one dominant resonant term.
This situation would be appropriate 
if the planet's eccentricity is very small in which case 
the $\delta_{k,0}$ term dominates.
The Hamiltonian (Equation \ref{hamgen}) including only this term 
\begin{eqnarray}
K(\Lambda, \psi; \Gamma,\gamma) 
   &  = & 
   a \Lambda^2 + b \Lambda + c \Gamma 
  \qquad  
\label{oneterm}
\\
& &  +  \delta_{k,0}    \Gamma^{k/2} 
      \cos{( \psi - k \varpi)}   \nonumber
\end{eqnarray}
It is convenient to perform a canonical transformation 
with generating function
\begin{displaymath}
F_2 = J_1\left({\psi\over k}- \varpi \right) + J_2\psi
\end{displaymath}
leading to new variables
\begin{eqnarray}
{J_1\over k}+ J_2 &=& \Lambda, \qquad \phi ={\psi \over  k}- \varpi \nonumber \\
J_1 &=& \Gamma, \qquad  \theta = \psi \nonumber
\end{eqnarray}
and new Hamiltonian
\begin{eqnarray}
K'(\Gamma,\phi;J_2,\psi) & = & a \left({\Gamma^2 \over  k^2} + J_2^2\right) 
         \qquad \nonumber \\
      &&   + \left({ 2a J_2\over k}+ {b \over k}+ c \right) \Gamma  
            + b J_2 \nonumber \\
&& + \delta_{k,0} \Gamma^{k/2} \cos(k \phi)   
\label{oneterm_g}
\end{eqnarray}
Note that $J_2$ is conserved and is small
for initial conditions near resonance with small initial 
eccentricity (or $\Gamma$).

Dropping constant terms and setting
$b' = (2 a J_2 + b) + kc$,
the Hamiltonian in Equation (\ref{oneterm_g})
\begin{eqnarray}
K'(\Gamma,\phi) 
   &  = & 
   {a \over k^2} \Gamma^2 + {b' \over k} \Gamma 
  \qquad \nonumber  \\
& &  +  \delta_{k,0}    \Gamma^{k/2} 
      \cos{(k \phi)}.  \nonumber
\end{eqnarray}
By rescaling momentum and time 
\begin{eqnarray}
\bar \Gamma  &=& \left|{ \delta_{k,0} k^2 \over a  }\right|^{-2/(4-k)} 
    \Gamma \nonumber \\
\tau &=&  |\delta_{k,0}|^{2/(4-k)}  
             \left|{ a\over k^2}\right|^{(2-k)/(4-k)}
                     t,
\label{rescale}
\end{eqnarray}
we can write this as 
\begin{equation}
\bar{K}(\bar \Gamma, \phi)   = {\bar\Gamma}^2 
        + \bar b \bar \Gamma   + (-1)^{k}{\bar\Gamma}^{k/2} \cos (k \phi)
\label{eqnKp}
\end{equation}
where
\begin{equation}
\bar b =  b' |\delta_{k,0}|^{-2/(4-k)}  
           \left|{ a\over k^2}\right|^{(k-2)/(4-k)} {\rm sign} (a).
\nonumber
\end{equation}
We relate the drift rate $\dot b$ from the migrating
planet to that of our scale-free system
\begin{equation}
\left|  {d \bar b \over d \tau} \right|= \left|{\dot b  \over k} \right|
           |\delta_{k,0}|^{-4/(4-k)}  
           \left|{ a\over k^2}\right|^{2(k-2)/(4-k)}.
\label{bardotb}
\end{equation}
The form of the Hamiltonian in Equation (\ref{eqnKp}) is 
(excepting for factors of $\sqrt{2}$)\footnote{
  By rescaling our momentum by a factor of
  $2^{k+2/(4-k)}$, the Hamiltonian becomes 
  $\bar\Gamma^2 + \bar b \Gamma 
  + (-1)^k 2^{(k+2)/2} {\bar \Gamma}^{k/2} \cos(k \phi)$ as by 
  \citet{M+D}.}
identical to 
that used to explore capture in the adiabatic limit 
(e.g., \citealt{henrard82,borderies84,malhotra90,M+D}).

\subsection{Capture probability as a function
of drift rate and initial momentum for first and second
order resonances} 
 
We ask the following question: 
{\it Above what drift rate ($d \bar b \over d \tau$) does the 
resonance fail to capture?} 
The system behaves adiabatically  when it takes longer than an oscillation
period for the system to pass through the resonance.
For low initial momentum, the width of the resonance is $\sim 1$,
and the period of oscillation is $\sim 1$.  
Consequently we expect
the system would evolve adiabatically when
$\left|{d \bar b \over d \tau}\right| \ll 1$.
To go beyond this limit and estimate the probability
of capture as a function of drift rate in the non-adiabatic
regime, we would need to find solutions to Hamilton's equation.
It is non-trivial to find solutions to the equations
of motion for Equation (\ref{eqnKp}) when $ \bar b$ is a function
of time and the system is not varying adiabatically 
(e.g., \citealt{friedland01}).
Consequently we have integrated
Hamilton's equations of motion numerically to explore
the non-adiabatic regime.
Once we numerically understand the behavior of the scale-free Hamiltonian
(Equation \ref{eqnKp}), we can make predictions for systems in the same
form using the factors of Equation (\ref{rescale}).

Our procedure for numerical integration is as follows.
Hamilton's equations for Equation (\ref{eqnKp})
are integrated using a conventional
Burlisch-Stoer numerical scheme.
The initial angle is chosen randomly.
We assume that ${\bar b}$ is proportional to time so that
only one parameter 
$d \bar b \over d\tau$ specifies the time dependence of the system.
To ensure that the particles were initially outside
of resonance we require the initial value of $|\bar b|$ to exceed 1.  
The parameter $\bar b$ was initially chosen to be $\sim -15$, well
outside the resonance.   
The system passes through resonance when $\bar b \sim 0$ so the timescale
until capture is $t_{capture} \sim b_{init} ({d b \over d \tau})^{-1}$.
The system is integrated at least twice the capture time.
Two sample integrations are shown in Figure \ref{fig:bfig}. 
Figure \ref{fig:bfig}a shows an integration illustrating a particle
that is captured into resonance and Figure \ref{fig:bfig}b shows
one where no capture takes place.

In the adiabatic limit,
the capture probability is 1 when the initial particle
eccentricity is smaller than a limiting value, $e_{lim}$, that 
depends on the order and width of the resonance
(e.g., \citealt{henrard82,borderies84,malhotra90,M+D}).
However when the drift is not adiabatic,
the capture probability could depend on 
the initial particle eccentricity (or momentum $\bar \Gamma$) 
even when it is below this limiting value.
Consequently we measured capture probability as a function of
both drift rate ($d \bar b \over d \tau$) and initial particle momentum.
For each value of drift rate and initial momentum,
we integrated
the system 100 times (each time with a different randomly chosen angle) 
to measure a capture probability. 

After the system captures, the momentum increases with time
(see Figure \ref{fig:bfig}a) and the resonant angle $\phi$ 
librates about a fixed value ($0$ or $\pi$) rather than circulating.  
If no capture takes place, the momentum
jumps as the system pass through resonance (see Fig. \ref{fig:bfig}b). 
When the momentum increases to a value exceeding the resonance
width (approximately 1 as we have rescaled the Hamiltonian) 
we identify the system as having captured into resonance.
We used a limiting value of $\Gamma = 5$ to identify captures.
In resonance the angle librates around
a fixed value.  The condition $\dot \phi \sim 0$ implies that 
${\partial K \over \partial \bar \Gamma} \sim 0$, and 
$\bar \Gamma \sim - \bar b/2$ in resonance.  
We use this condition to ensure that we integrate Hamilton's equations 
long enough that the momentum crosses our limiting momentum value
when the resonance captures.

Figure \ref{fig:simplek1} and \ref{fig:simplek2} show
capture probabilities that we have measured numerically for first and second
order resonances ($k=1$ and $2$).  
In the adiabatic limit for 
$\bar \Gamma(t_0)$ below 3/2 for $k=1$, or $1/8$ for $k=2$, the capture 
probability 
is one.\footnote{Our momentum is half or 1/4 times that of \cite{M+D} 
for $k=1$ and $k=2$ respectively.   \cite{M+D} list
critical momentum values of 3 and $1/2$.
These critical values are used to find the maximum particle
eccentricity, $e_{lim}$, that ensures capture in the adiabatic limit 
\citep{henrard82,borderies84,malhotra90}.}
Above these limiting initial momentum values the capture 
probability is less than 1 in the adiabatic limit.
For $\bar \Gamma(t_0) = 2.3$, shown as stars 
in Figure \ref{fig:simplek1}, above the limiting 
value of $\bar \Gamma_{0,lim} = 1.5$, we see that the capture probability never
reaches 1.  At lower drift rates
the capture probability approaches a constant value for this
initial condition, consistent with the prediction in the
adiabatic limit.  The same behavior is seen for $k=2$
with initial momentum $\bar\Gamma(t_0) = 1$ (shown as stars
in Figure \ref{fig:simplek2}). This momentum is eight times the limiting 
value which ensures capture in the adiabatic limit.

To quantify the width of the probability function,
we have fit a function to the capture probability 
$p(u) = 0.5\left( 1 - \tanh \left({u-u_{1/2}\over w}\right) \right)$ 
as a function of 
$u = \log_{10}\left|{db \over d \tau}\right|$. Here $u_{1/2}$ is the log
of the drift rate at which the capture probability is 1/2, and $w$
describes the width of the drop.  For large $w$, the slope
is shallow, for small $w$ the drop is a steep function of the drift rate.
The drift rates at which the capture probability is half and a quarter,
and the widths of the probability functions
are shown as a function of initial momentum in Figures 
\ref{fig:gamk1} and \ref{fig:gamk2} for first and second order
resonances, respectively. 
For initial momentum sufficiently low, (e.g., $10^{-2}$ for $k=1$
and $10^{-6}$ for $k=2$)  the drift rate at which the probability
is half approaches a limiting value. 
The steepness of the transition between 100\% capture and
0\% capture is narrower in its range of drift rates at lower values
of initial momentum.  
The lower the initial momentum, the sharper the transition
between a capture probability of 1 and zero.
A sharp transition is reached at a lower initial momentum for
$k=2$ than in the $k=1$ case.
For initial momentum near $\bar \Gamma_{0,lim}$, 
the limiting value ensuring capture in
the adiabatic limit, there is a regime or a range of drift
rates where the capture probability is intermediate.
In other words, for $\bar \Gamma(t_0) \sim 1$ the widths $w \sim 1$.

We now consider the situation where the transition
between a probability of 1 and 0 is sharp.
This is true for initial momentum 
$\bar \Gamma (t_0) \lesssim 10^{-2}$ and $10^{-6}$
for $k=1$ and for $k=2$ respectively. 
For these initial momenta we measure the critical drift rate where
the transition takes place.
From our numerical integrations,
the dynamical system fails to capture for
drift rates faster than
\begin{eqnarray}
\left|{d \bar b \over d \tau}\right|_{crit}   \sim & 2.0 
        \qquad & {\rm for} ~ k=1  \nonumber \\
               \sim & 0.25 
        \qquad & {\rm for} ~ k=2.
\label{eqndb}
\end{eqnarray}

For first order resonances the drift rate for a capture probability of
1/2 is not strongly dependent on the initial momentum as long
as this lies below $\bar \Gamma_{0,lim}$.
However for second order resonances 
when the initial momentum $\bar \Gamma(t_0) \sim 1$,
the capture probability of half occurs
at a drift rate that is about 10 times that at low initial momentum.
We have approximated the dependence of the half probability drift rate
on the initial momentum
with the following function (shown as a dotted line 
in Figure \ref{fig:gamk2})
\begin{equation}
\left|{d \bar b\over d \tau}\right|_{1/2}
\sim 0.25\left( 1 + {\bar \Gamma (t_0) \over 3\times 10^{-5} }  \right)^{0.25}.
\label{eqnhalf}
\end{equation}
The power, 0.25, is not necessarily theoretically meaningful.
This function is a reasonable match to the measured points 
of Figure \ref{fig:gamk2} for initial
momentum $\bar \Gamma (t_0) \lesssim 1$. 
For higher initial momentum, $\bar \Gamma (t_0) \gtrsim 4$, the probability of
capture never exceeds 1/2.  This initial momentum (4) exceeds 
$\bar\Gamma_{0,lim}$ by a factor of 32.

In the adiabatic limit,
the probability of capture drops as a function of increasing
momentum (or eccentricity)
when the initial momentum is above $\bar \Gamma_{0,lim}$.
However the probability drops faster for first order resonances
than for second order resonances \citep{hahn99}.
In the adiabatic limit, the probability of capture for a first
order resonance drops to
1/2 for $\bar \Gamma (t_0) \sim 2.3$ (less than twice the limiting 
value of 1.5), whereas  for second order resonances the probability
of capture drops to 1/2 for $\bar \Gamma (t_0) \sim 4$ or 32 times
the limiting value.

Using the critical drift rates for ${d \bar b\over d \tau}$ 
(equation \ref{eqndb})
we can invert Equation (\ref{bardotb}) to determine
which resonances can capture at a particular drift rate.
We find that resonances are likely to capture for $\dot b$ slower than
the critical rates
\begin{eqnarray}
|\dot b_{crit}|  \sim & 2 |\delta_{1,0}|^{4/3} |a|^{2/3} 
        \qquad & {\rm for} ~ k=1  \nonumber \\
  \qquad \qquad \sim & 0.5 \delta_{2,0}^{2} 
        \qquad & {\rm for} ~ k=2. 
\end{eqnarray}
Using Equation (\ref{eqnab}) to replace $\dot b$ with the planet's
mean motion
\begin{eqnarray}
|\dot n_{p,crit}| \sim & 2(j-1) |\delta_{1,0}|^{4/3} |a|^{2/3} 
        \qquad & {\rm for} ~ k=1  \nonumber \\
  \qquad \qquad \sim & 0.5(j-2) \delta_{2,0}^{2} 
        \qquad & {\rm for} ~ k=2.
\label{eqnnpcrit}
\end{eqnarray}
For second order resonances,
the rate given above can be modified by
the function given in Equation (\ref{eqnhalf}) to estimate
the rate at which the probability is 1/2 
as a function of initial momentum;
\begin{equation}
|\dot n_{p,1/2}| \sim  0.5(j-2) \delta_{2,0}^{2} 
      \left( 1 + {   \alpha^{-1/2} e_0^2 a\over 2.4 \times 10^{-4} \delta_{2,0} } 
        \right)^{0.25}
\label{eqnnphalf}
\end{equation}
where $e_0$ is the initial particle eccentricity.
This is expression is valid for initial particle eccentricities smaller
than $e_0 \lesssim 10 e_{lim}$ where $e_{lim}$ is the eccentricity
limit ensuring capture in the adiabatic limit.  
An expression for $e_{lim}$ is given in the appendix.
The factor of 10 comes from the range covered
by the function shown in Equation (\ref{eqnhalf}).
The curve shown in Figure \ref{fig:gamk2} 
is a reasonable match up to $\Gamma (t_0) \sim 1$
which is approximately $10 \bar \Gamma_{0,lim}$.

The above relations (Equations \ref{eqnnpcrit} and \ref{eqnnphalf}) 
allow us to estimate the likelihood
of resonance capture in different astronomical settings.
The strength of
the perturbative cosine terms ($\delta_{k,p}$) are proportional to the planet's
mass or $\mu$,  however the critical drift speed
depends on $\mu^{4/3}$ for $k=1$ and on $\mu^{2}$ for $k=2$.
We see that the critical drift rates for capture are strong functions
of the planet mass and this is particularly true for
the second order resonances.  
Slower drift rates are required to allow resonant capture
 for lower mass planets.

The dependence of critical drift rate on planet
mass provides a qualitative explanation for some features
of numerical simulations which start with particles
in initially low eccentricity orbits.   
The above relation predicts that only more slowly drifting
particles will be able to capture into higher
order resonances.  We can understand why 
the 5:3 resonance requires slower migration rates than the 2:1 and 3:2
resonances to capture in the simulations of \citet{wyatt03}.
Although we have integrated a time dependent Hamiltonian 
system, we can
expect similarities between this system and the slowly
drifting non-conservative systems.
Simulations of dust drifting inward via dissipative forces tend to show
that large dust particles are captured into higher order
resonances than smaller particles (e.g., \citealt{marzari94,liou99}).
This follows since small particles drift faster than larger ones
and the higher order resonances require slower drift rates
to capture. 

Based on results of simulations of Neptune's migration
\citet{ida00} proposed that the critical drift rate 
depended on planet mass to the $4/3$ power.
They restricted their study to $k=1$ resonances so their
prediction is consistent with our previous equation.
This power dependence was confirmed with analytical work by \citet{friedland01},
also for the $k=1$ resonance, and numerical work by \citet{wyatt03}.
We confirm the steeper dependence on planet mass
of the 5:3 resonance capture probability
measured numerically by \citet{wyatt03} and specifically predict
that the critical drift rate is $\propto \mu^2$ for 
second order resonances.  
The relation for the critical drift rate (Equation \ref{eqnnpcrit})
is both consistent with and more general than the scaling found
by these previous studies.  Because we have related the critical
drift rate (via scaling) to the resonance strengths, the formulation
given here can be applied to any first or second order mean motion 
resonance.

Here we have also found that the probability of capture
when the drift is not adiabatic 
is a non-trivial function of initial particle
eccentricity.  The transition between
a probability of 1 and 0 becomes smoother (covering a larger
range of drift rates) as the initial momentum approaches 
the minimum value ensuring capture in the adiabatic limit
(see Figures \ref{fig:simplek1}, \ref{fig:simplek2}, 
\ref{fig:gamk1}, and \ref{fig:gamk2}).
For first order resonances, the midpoint drift rate
(corresponding to a probability of capture of 1/2) 
does not significantly depend
on the initial particle eccentricity. However for second
order resonances the midpoint is at a higher drift rate
when the initial momentum or particle eccentricity is higher.  
The increase in drift rate with initial particle eccentricity
allowing capture for second order resonances
was described previously by \citet{hahn05}.
However we do not predict the same dependence on resonance width
and planet mass. This is because
we have restricted our study to 
initial particle  eccentricity near or below $e_{lim}$,
and \citet{hahn05} considered 
initial particle eccentricity exceeding $e_{lim}$.

Our estimate for the
critical drift rate above (Equation \ref{eqnnpcrit})
is appropriate for a wide range of initial particle eccentricities
for first order resonances (as long as they are below the
limiting value, $e_{lim}$).  The half probability 
drift rate's dependence on the initial particle eccentricity can
be estimated for second order resonances
using Equation (\ref{eqnnphalf}) when the initial
particle eccentricity
is lower than $\sim 10$ times $e_{lim}$. 
The framework we provide here can we used to estimate the half
probability drift rate for any second order resonance.

Numerical studies report intermediate probabilities
for capture into first order resonances 
from simulations \citep{ida00,quillenholman00,chiang02,wyatt03}.  
By intermediate, we mean not close to zero or 1, 
or at a $\sim 50\%$ level.
Here we have found that the dynamical system 
described by Equation (\ref{eqnKp}) for $k=1$ with only a single resonance
term has a limited range of drift rates where
the capture probability is intermediate, unless the initial
particle momentum $\bar \Gamma$ is of order 1.  This regime
corresponds
to an initial particle eccentricity within a factor of a few 
of $e_{lim}$, the limiting value ensuring capture in the adiabatic limit.
The limiting eccentricity depends on the resonance strength,
and planet mass to the power $k/(4-k)$ (using the square
root of the scale-free momentum in Equation (\ref{rescale});
\citep{malhotra90,M+D}.  For weaker resonances, 
the initial particle eccentricity limit is more restrictive.
It is possible that some of the numerical simulations are effectively
in the regime of intermediate capture for certain resonances due
to their initial particle eccentricity distribution.  
We return to this issue in later sections as
we identify other regimes of intermediate capture probability for
first order resonances.

Though the limiting eccentricity is smaller for second order resonances,
the probability of capture drops more slowly in the
adiabatic limit as a function of initial particle 
eccentricity at values above $e_{lim}$.
Furthermore since the half probability drift rate increases 
with initial particle eccentricity (Equation \ref{eqnnphalf}), 
higher eccentricity particles
can be captured at higher drift rates than lower
eccentricity particles (as pointed
out by \citealt{hahn05}).  This also implies that
the second order resonances have a larger
regime in both range of initial eccentricity and drift
rate where the probability of capture is intermediate.

\section{The role of an additional resonance term}

We now consider the differences in the dynamics of
capture caused by the addition of a secondary
resonant term for a first order resonance.
We rescale the momenta and time for 
Hamiltonian given in Equation (\ref{hamk1})
for $k=1$
according to the Equations(\ref{rescale}).
This gives us unitless momenta and time
\begin{eqnarray}
\bar K_1 (\bar \Lambda, \psi; \bar \Gamma, \gamma)
& = & \bar\Lambda^2 + \bar b \bar \Lambda + \bar c \bar \Gamma    
\label{eqnvarne}
\\
\qquad & & - \bar \Gamma^{1/2} \cos(\psi - \varpi)
           + \bar \epsilon \cos(\psi - \varpi_p) \nonumber
\end{eqnarray}
where
\begin{eqnarray}
\bar \epsilon  &=&|\delta_{1,1}||\delta_{1,0}|^{-4/3} |a|^{1/3} \nonumber \\
\bar  c        &=& c |\delta_{1,0}|^{-2/3} |a|^{-1/3} { \rm sign} (a).
\label{eqnec}
\end{eqnarray}
The coefficients that can be adjusted are the width of the
second resonance compared to the first (set by $\bar \epsilon$)
and the separation between the two resonances (set by $\bar c$).
Variation in $\dot \varpi_p$ can be absorbed into our 
coefficient $\bar c$.
The term proportional to $\cos(\psi - \varpi)$ is often
called the $e$-resonance since $\Gamma^{1/2} \propto e$.
The other term can be called an $e'$-resonance 
or a corotation resonance since it
does not depend on the particle's longitude of perihelion or $\varpi$.

Because the corotation resonance does not depend on $\bar \Gamma$, it
does not grow in volume
as the planet migrates.   Had we allowed $\alpha$ to depend upon
time, the resonance width would grow slightly but not significantly
as the planet migrates.  Because the resonance volume in
phase space does not grow
as the planet migrates, this resonance
should not capture particles \citep{yoder79}.   
However when this resonance
overlaps the other, the system can exhibit large scale chaotic behavior
\citep{holman96}.
Hence the coupling of the two resonant terms may influence the probability
of capture into the $e$-resonance.

We now ask: {\it what is the capture probability of the above Hamiltonian
as a function of drift rate, ${d \bar b \over d \tau}$,  and
secondary perturbation strength, $\bar \epsilon$?
}
To answer this question we numerically integrate Equation (\ref{eqnvarne}),
for different parameters $d \bar b \over d\tau$, 
and $\bar \epsilon$.   In this section we work in the limit
of low initial particle momentum.  This ensures that
the probability of capture is 1 in the adiabatic limit and that
the transition between capture and no capture would be a sharp
transition of drift rate in the case of a single resonance.

Our procedure for numerical integration is the same
as described in the previous section.
Initial angles are randomly chosen.  Initial momenta 
($\bar \Lambda, \bar \Gamma$) are set to small values to 
ensure a sharp transition when $\bar\epsilon=0$ and the
initial momentum are small.
The parameter $\bar b$ is initially chosen to be $\sim -15$.
For each value of ${d \bar b \over d \tau}$ and $\bar \epsilon$ we integrated
the system 100 times to estimate a resonant capture probability. 
After the system captures, the momenta variables increase with time
and the resonant angle $\phi = \psi - \varpi$ 
librates about a fixed value.  If no capture takes place, the momenta
jump as the system pass through resonance.  

The coupled two dimensional system exhibits different 
dynamics than the one-dimensional system considered in the
previous section.
For example, the resonance can capture for a short period of
time, a trajectory we refer to as a temporary capture.
An example of a simulation that illustrates
a temporary capture is shown in Figure \ref{fig:partcap}.
We find that temporary captures tend to occur
for larger values of $\bar \epsilon$ and drift rate.
\citet{quillen_cap} previously showed
that temporary capture
was exhibited by overlapped resonances using a similar 
drifting Hamiltonian model.

During a temporary capture, the momenta increase. This also happens
if the particle is captured.
As in the previous section, we identify a capture if the momentum
at the end of the integration exceeds a value of 5.  However we then
reclassify the integration as a temporary capture 
if the momenta
lie below that expected from a particle still
in resonance. These two situations can be differentiated
because a particle in resonance has momentum proportional
to the time since capture.   
Temporary captures are excluded when we calculate the capture probability. 
However,
had we integrated the systems longer, it is possible that
a particle identified on a short timescale
as captured would later drop out of resonance.
In other words the precise fraction of captures 
is dependent on the timescale over which we have integrated these systems
and the value of momentum that we have used as a limit to identify captures.
This makes our capture probability numerical measurements uncertain
primarily at high values of drift rate 
$|{d\bar b \over d \tau}|\gtrsim 1$, and large secondary
perturbation strength $\bar \epsilon \gtrsim 1$, the regime
where we have found temporary captures to be more common.

In Fig. \ref{fig:a1n} we show a contour plot of the resonant
capture probability for $k=1$, initial $\bar \Gamma(t_0) = 10^{-4}$ 
and resonance separation $\bar c=0$.
For high values of the drift rate and low values of the
secondary perturbation, $\bar \epsilon$,  (on lower right
in these contour plots)
the transition between capture and no capture is sharp
and happens at the critical drift rate measured in the previous
section.  However for lower drift rates and higher values of 
$\bar \epsilon$ the resonance fails to capture for $\bar \epsilon \gtrsim 1$.
For low drift rates, the transition between capture and no capture
is also sharp, but is a function of $\bar \epsilon$ instead of drift rate.
There are two regimes, that where the drift is so fast
that it fails to capture, and that where the corotation resonance is
so large that it prevents capture.

For first order resonances we consider the possibility that 
$\bar \epsilon$ could be of order 1.
Since $\bar \epsilon \propto |\delta_{1,0}|^{-4/3}$, 
$|\delta_{1,0}|\propto \mu$ and $|\delta_{1,0}| \propto \mu e_p$,
the coefficient $\bar \epsilon\propto \mu^{-1/3} e_p$.  For small planet mass
$\mu$ we see that the rescaled secondary perturbation strength 
could be high even at moderate planet eccentricity $e_p$.
The coefficient $\bar \epsilon$ could
be of order 1 particularly if the planet eccentricity is moderate.

Why is it that for $\bar \epsilon \gtrsim 1$ 
the $e$-resonance fails to capture particles?
A possible explanation is that an increase in $\bar \Gamma$
caused by the corotation reduces the capture
probability, in the same way that an increase in the initial
momentum value does. We would expect that an increase
of initial momentum of size $\sim 1$ caused by
the corotation resonance when $\bar \epsilon \sim 1$ would strongly reduce
the capture probability.  This is consistent with the limiting
value of $\bar \epsilon \lesssim 1$ for capture.  This 
qualitative explanation is also consistent 
with the lack of dependence of the critical value of $\bar \epsilon$
on drift rate  (see Figure \ref{fig:a1n}) at drift rates 
$\left|{d \bar b \over d \tau}\right|< 1$.
We suspect that the corotation resonance prevents capture
into the $e$-resonance
because the corotation resonance raises the particle eccentricity
during the resonance encounter.
If this were true, then we would expect that the capture
probability would be influenced by the resonance separation.
Up to this point we have only considered resonances 
with separation $\bar c=0$.

\subsubsection{Separated first order resonances}

To further explore the role of multiple resonant terms, we consider 
the situation when the resonance separation is non zero 
($\bar c$ not small).  
The order that the subresonances are encountered 
as the system drifts
can be determined by considering the two resonant angles
$\psi - \varpi$ and $\psi - \varpi_p$.
The time derivative
$\dot \psi - \dot \varpi =  j n - (j-1) n_p - \dot \varpi$.
As the planet migrates outward $n_p$ drops.  For $\dot \varpi > \dot \varpi_p$
the time derivative of the corotation resonant angle crosses zero  first.
For $\bar c > 0$ (corresponding to $c < 0$ and a positive
precession rate for the longitude of periapse), 
the corotation resonance
is encountered first by a particle exterior
to a planet as the planet migrates outward.
This is what is expected for external resonances in a single
planet system where the precession rate of the planet's longitude of periapse
$\dot \varpi_p =0$ and $\dot \varpi >0$ due
to secular precession induced by the planet.

We consider which resonance is encountered first for other drifting
systems.
Dust particles migrating inward and exterior to a planet 
would encounter the resonances
in the same order, corotation resonance first 
as long as $\dot \varpi > \dot \varpi_p$.   
Here $n$ is increasing, whereas for the planet
migrating outward $n_p$ was decreasing. 
A particle located internal to a planet that is
migrating inward would also encounter the resonances in 
the same order when $\dot \varpi > \dot \varpi_p$;   
(in this case one considers
$(j+1)n_p - j n - \dot \varpi$ and  
$(j+1)n_p - j n - \dot \varpi_p$  with $n_p$ increasing).

We can compare the width of the corotation resonance to the separation
between them.
For the $e$-resonance, the libration width in $\bar \Lambda$ depends on the
particle eccentricity or $\bar \Gamma$.  However since the corotation
term does not depend on $\bar \Gamma$, the
corotation resonance width can more easily be estimated as
$\Delta \bar \Lambda \sim \sqrt{ \bar \epsilon}$.
For 
\begin{displaymath}
|\bar  c| \lesssim \sqrt{\bar \epsilon},
\end{displaymath}
the two resonances must overlap.
Since $\bar c \propto \mu$ and 
$\sqrt{\bar \epsilon} \propto e_p^{1/2} \mu^{-1/6}$,
strong resonances are likely to overlap for small planet
masses unless the planet eccentricity is extremely small.
Because the two resonant terms differ in sign,
they have fixed points at different angles, and they 
are expected to interfere when overlapped 
even only slightly separated.

We have found that
a nonzero value of $\bar c$ does change the probability of capture.
Figure \ref{fig:a2n} shows numerical  measurements similar
to that of Figure \ref{fig:a1n} but for $\bar c = \pm 0.1$.
For $\bar c <0$ (Figure \ref{fig:a2n}a) 
the corotation resonance is encountered
after the $e$-resonance. 
The onset of the corotation resonance can knock the particle
out of resonance, following capture into the $e$-resonance.
For $\bar c >0$ the corotation resonance is encountered first
as the planet migrates (Figure \ref{fig:a2n}b).
The capture probabilities are primarily modified at low drift
rates where higher values of 
$\bar \epsilon$ are required to reduce the probability of capture.
It is not obvious why this is the
case.   From individual integrations we note that the frequency 
$\bar c$ sets an oscillation period that is longer
for smaller values of $\bar c$.   \citet{holman96} found
that $\bar c$ sets the Lyapunov time of the resonance.
So for an overlapped system
we might expect more highly chaotic behavior for larger
values of $\bar c$, particularly at low drift rates.  Oddly
higher values of $\bar c$ at low drift rates
seem to stabilize the system, requiring higher values
of the corotation resonance strength to kick the particle out of the 
$e$-resonance.  

For $\bar c <0$ (as shown in Figure \ref{fig:a2n}b) the corotation
resonance is encountered after the $e$-resonance, consequently  
the corotation resonance is encountered after the $e$-resonance captures
a particle.    If the corotation resonance is strong it can
knock the particle out of resonance.  For larger separations, $\bar c \sim 1$,
temporary captures are frequent at large $\bar \epsilon$.
The extended low probability contours on the top end of Figure \ref{fig:a2n}b 
are in part due to temporary captures.

\subsection{Second order resonances}

When the resonance is second order it contains three subterms 
(see Equation \ref{hamgen}).
The first term $\propto \Gamma \cos(\psi - 2 \varpi)$ and can 
be called (e.g., \citealt{M+D}) 
the $e^2$-resonance since $\Gamma \propto e^2$ at low eccentricity.  
The second term 
 $\propto e_p \Gamma^{1/2}  \cos(\psi - \varpi - \varpi_p)$ and 
can be called an $ee'$-resonance. The third term
 $\propto e_p^2  \cos(\psi -2 \varpi_p)$ can
be called a corotation or $e'^2$ resonance.
Since the corotation term does not depend on $\Gamma$,
its volume in phase space does not grow as the planet drifts
and it should not be able to capture particles. 
However, as
was true for the first order resonance, this resonant term
can prevent the other resonant terms from capturing particles.
When the corotation term is not large,
both the $e^2$- and $ee'$-resonances can capture particles.
Since it is $ \propto \Gamma^{1/2}$, 
the $ee'$ term behaves like a first order resonant term,
whereas the $e^2$ term, $\propto \Gamma$, is a second order term.
In the previous section we found that first order resonances
captured at a somewhat higher drift rate than the second order
term and did not require as low initial momenta to exhibit a sharp
transition between capture and no capture.  The critical
drift rate for first order terms is $\propto \mu^{4/3}$ and for
second order terms $\propto \mu^2$, a much steeper function of planet mass.  
Consequently it is possible that the $ee'$ resonant term will
capture particles and the $e^2$-resonance will not capture particles
even when the planet eccentricity is low. 

We first consider the situation where the $e^2$-resonance
is dominant.
Taking $k=2$ terms from Equation (\ref{hamgen}), 
we rescale the Hamiltonian as follows,
\begin{eqnarray}
\bar K_\xi \bar \Lambda, \psi; \bar \Gamma, \gamma)
& = & \bar\Lambda^2 + \bar b_\xi \bar \Lambda + \bar c_\xi \bar \Gamma    
           + \bar \Gamma \cos(\psi - 2 \varpi)
\label{eqnvarnexi}
\\
\qquad & & 
           - \bar \xi  \bar \Gamma^{1/2} \cos(\psi - \varpi - \varpi_p)
           + \bar \epsilon_\xi \cos(\psi - 2 \varpi_p) \nonumber
\end{eqnarray}
where
\begin{eqnarray}
\bar \Gamma    &=&  \left|{\delta_{2,0}\over a}\right|^{-1}
                     \Gamma \nonumber \\
\tau           &=&  |\delta_{2,0}|  t \nonumber \\
\bar  c_\xi    &=& c |\delta_{2,0}|^{-1} {\rm sign}(a) \nonumber \\
\bar  b_\xi    &=& b |\delta_{2,0}|^{-1} {\rm sign}(a) \nonumber \\
\bar \xi       &=& |\delta_{2,1}||\delta_{2,0}|^{-3/2}  |a|^{1/2}
    \nonumber \\
\bar \epsilon_\xi  &=& |\delta_{2,2}||\delta_{2,0}|^{-2} |a|.
   \nonumber 
\label{eqnxi}
\end{eqnarray}

The coefficients that can be adjusted are the strength of the
$ee'$-resonance (set by $\bar \xi$),
the strength of the corotation resonance (set by $\bar \epsilon_\xi$), 
and the separation between the resonances (set by $\bar c_\xi$).
Since the precession rate $ c \propto \mu$ and 
the perturbation strength $\delta_{2,0} \propto \mu$ we find
that the resonance separation ($\bar c$) does not depend on the planet mass.
This implies that the subterms could be
well separated. This is different than the first order
resonances that have $\bar c \propto \mu^{1/3}$,
which implies that the resonance subterms are often overlapped. 

The $ee'$-resonance strength
$\bar \xi \propto \mu^{-1/2} e_p$.  Since this depends
on a negative power of $\mu$, at low planet
masses and at high planet eccentricities it is possible that $\bar \xi > 1$.
In this case the $ee'$-resonance could be dominant,
and we would rescale momentum and time as we did
for a first order resonance.
In this case we could work with the Hamiltonian
(equivalent to the previous one except for the rescaling)
\begin{eqnarray}
\bar K_\chi (\bar \Lambda, \psi; \bar \Gamma, \gamma)
& = & \bar\Lambda^2 + \bar b_\chi \bar \Lambda + \bar c_\chi \bar \Gamma    
\label{eqnvarnechi} \\
\qquad & & + \bar \Gamma^{1/2} \cos(\psi - \varpi - \varpi_p) \nonumber \\
\qquad & & 
           - \bar \chi      \bar \Gamma \cos(\psi - 2 \varpi)
           + \bar \epsilon_\chi  \cos(\psi - 2 \varpi_p) \nonumber
\end{eqnarray}
where
\begin{eqnarray}
\bar \Gamma    &=&  \left|{\delta_{2,1}\over a}\right|^{-2/3}
                     \Gamma \nonumber \\
\tau           &=&  |\delta_{2,1}|^{2/3} |a|^{1/3}   t \nonumber \\
\bar  b_\chi   &=& b |\delta_{2,1}|^{-2/3} |a|^{-1/3} {\rm sign}(a) \nonumber \\
\bar  c_\chi   &=& c |\delta_{2,1}|^{-2/3} |a|^{-1/3} {\rm sign}(a) \nonumber \\
\bar \chi      &=& |\delta_{2,0}| |\delta_{2,1}|^{-2/3}  |a|^{-1/3}
    \nonumber \\
\bar \epsilon_\chi  
               &=& |\delta_{2,2}||\delta_{2,1}|^{-4/3} |a|^{1/3}.
\label{eqnchi}
\end{eqnarray}

A comparison between Equation (\ref{eqnxi}) and Equation (\ref{eqnchi})
shows that 
\begin{equation} 
\bar \xi^2 = \bar \chi^{-3}
\end{equation} 
We suspect that the $e^2$-resonance is more likely to 
capture when $\bar \xi$ reaches a transitional value that we denote 
$ \bar\xi_{trans}$.
The $ee'$-resonance may be more important 
when $\bar \chi \lesssim \bar\xi_{trans}^{-2/3}$.
If $\bar \xi \lesssim \bar \xi_{trans}$ then we expect second order behavior 
($\dot n_{p,crit} \propto \mu^{2}$) otherwise we expect
first order behavior
($\dot n_{p,crit} \propto \mu^{4/3}$).
We note that the order the resonances are encountered is also important.
Here we numerically measure a capture probability that does not
specify which resonance captures.
If the drift rate is sufficiently slow that the $e^2$-resonance
captures and this resonance is reached first then this resonance
will dominate the capture probability.

As we did for the $k=1$ resonances we have measured the capture
probability for a range of coefficients and drift rates.
Figure \ref{fig:xichi}a shows 
the capture probability for 
the Hamiltonian of Equation (\ref{eqnvarnexi}) with varying drift rate,
and $ee'$-resonance perturbation strength,
$\bar\xi$, and with no corotation term; $\bar \epsilon_\xi=0$.
The resonances are not separated; $\bar c_\xi =0$.
On the lower left side of this plot, for weak $\bar \xi$ we see
capture behavior consistent with the pure second order ($k=2$)
system discussed in the last section with only one resonant term.
With the variables defined in this section (which differ
by a factor of $k=2$ from those defined in Equation \ref{rescale})
$\left|{d \bar b \over d \tau }\right|_{crit} = 0.5$.
The capture probability ceases to depend on $\bar \xi$ for
$\bar \xi \lesssim 10^{-2}$.
Consequently we can estimate a transition value 
\begin{equation}
\bar \xi_{trans}\sim 10^{-2} 
\end{equation}
valid for low initial particle momentum.
For $\bar \xi \gtrsim \xi_{trans}$, the transition 
between capture and no capture occurs at faster drift rates
and the transition drift rate is a function of $\bar\xi$.
For $\bar \xi > \bar \xi_{trans}$ the system behaves like a first
order system, and since the perturbation strength depends on 
$\bar \xi$ we expect the transition drift rate to depend
on $\bar \xi^{4/3}$. This is consistent with the trend shown
on the upper right in Figure \ref{fig:xichi}a.
For large $\bar\xi$ and at high drift rates the system can fail
to capture into the $e^2$-resonance but can be captured into
the $ee'$-resonance.
In short, for $\bar \xi \lesssim \bar \xi_{trans}$  the system behaves
like a second order system and tends to capture into the
$e^2$-resonance, however for $\bar \xi \gtrsim \bar \xi_{trans}$
the highly overlapped 
system behaves like a first order system and tends to capture
into the $ee'$-resonance.  

Figure \ref{fig:xichi}b shows the capture probability for 
the Hamiltonian of Equation (\ref{eqnvarnechi}) with
varying drift rate, varying $e^2$-resonance perturbation strength, 
$\bar \chi$ and with no corotation term;  $\bar \epsilon_\chi =0$.
The resonances are not separated; $\bar c_\chi =0$.
This figure extends Figure \ref{fig:xichi}a to $\bar \xi > 1$ 
because $\bar \chi = \bar \xi^{-2/3}$.
At small $\bar \chi$ the system critical drift rate
is independent of $\bar \chi$ and consistent with
that measured in section 3 for a first order resonance with   
$\left|{d \bar b \over d \tau }\right|_{crit} = 2$.
At this limit the system behaves like a first order resonance.  
The second order regime is not fully reached until  
$\bar \xi \sim \bar \xi_{trans}$ which would correspond to 
$\bar \chi >20$.   However, for large values of $\bar \xi$
we see that the critical drift rate does begin to increases, consistent with
a dependence of critical drift rate on $ \bar \chi$ (the second
order term dominates).
In Figure \ref{fig:xichi}b where 
$ \bar \chi \sim 0.5$ we see an extended
region of drift rates corresponding to a regime of intermediate
capture probability.  
We have inspected individual integrations from this region 
and seen temporary captures and widely varying or chaotic 
trajectories.
The intermediate capture probability measured
is probably due to 
interference between the two similarly sized and overlapped
$e^2$- and $ee'$-resonances.

Figures \ref{fig:xichi} shows integrations done for low initial
particle eccentricity (low $\bar \Gamma (t_0)$). 
We discuss what we expect would happen for higher initial particle
momentum.
The second order resonance is more strongly affected by
the initial particle momentum.   For larger $\bar \Gamma (t_0)$
the drop in probability at the bottom right on Figure \ref{fig:xichi}a
would occur at a faster drift rate and the drop would be smoother.
The contours would broaden and shift to the right, primarily
on the bottom of Figure \ref{fig:xichi}a.  The resonance behaves
like a second order resonance at the top right of Figure \ref{fig:xichi}b.
So we expect a similar broadening and shifting to the right of the contours
at the top of Figure \ref{fig:xichi}b.

We now investigate the role of the corotation or $e'^2$ resonance term.
When the $e^2$ term is not important 
(as we have found $\bar \xi \gtrsim \bar \xi_{trans}$), the system is
identical to that studied in the previous section 
for $k=1$ with two terms (Equation \ref{eqnvarne}).
However we can also study the system for Equation \ref{eqnvarnexi}
as a function of $\bar \epsilon_\xi $ and with $\bar\xi =0$ so the corotation
term is strong but the $ee'$-resonance is weak.  
Figure \ref{fig:b1n} shows
the capture probability for 
the Hamiltonian of \ref{eqnvarnexi} with varying $\bar\epsilon_\xi$ 
drift rate, $d \bar b \over d \tau$, and with $\bar \xi = \bar c_\xi =0$.
We find here that even moderate values of the corotation
resonance strength $\bar \epsilon \sim 0.1$
can significantly reduce the capture probability.
The corotation resonance can reduce the probability of 
capture for both first and second order resonances.

\subsubsection{Separated second order resonances}

As was true for the first order resonances we expect the
capture probability to depend on the resonant term separations
and order that the resonances are encountered.
Figure \ref{fig:xic} shows the effect of changing 
the resonance separation $\bar c_\xi$ 
when the $e^2$-resonance dominates.
Figure \ref{fig:xic}a shows the case with widely separated resonances
when the $e^2$-resonance is encountered first.  We see that the transition
value of $\xi_{trans}$ is higher than when there is no separation ($\bar c=0$).
When the $e^2$-resonance is encountered afterward the transition value
of $\xi_{trans}$ is lower.  The $ee'$-resonance interferes with
the capture into the $e^2$-resonance to a higher degree when
this resonance is encountered earlier.
Figure \ref{fig:xic} shows the effect of changing 
the resonance separation $\bar c_\chi$ 
when the $ee'$-resonance dominates.  We find that the region of 
intermediate capture probability at 
$\bar  \chi \sim 0.5$ is smaller when the resonances
are separated than when $\bar c =0$ (Figure \ref{fig:xichi}b).


\section{Applications}

In this section we apply what we have learned above to two 
systems involving capture into the 2:1 resonance.  When the 2:1 resonance
is exterior (and capture particles as a planet migrates outward) 
the $e$-resonance strength is reduced because of
the indirect term.  This reduces the critical
migration rate compared to that for other first order resonances.  
Because the corotation resonance is not affected by the indirect term, it
is comparably strong.  Consequently even a small planet eccentricity
can reduce the capture probability.  When the 2:1 resonance is
interior (and can capture for a planet migrating inward)
the indirect term reduces the strength of the corotation resonance instead
of the $e$-resonance.
In this case the 2:1 resonance is strong and can capture at fairly
high migration rates.  We consider two situations, the capture
of twotinos into the 2:1 resonance by Neptune migrating outward,
and the capture of an inner extra solar planet into the 2:1 or 3:1
resonance by an inward migrating planet exterior to it.

\subsection{The capture of twotinos in the Kuiper Belt via Neptune's migration}

In this section we consider the capture of Kuiper belt objects
into the 2:1 resonance  by an outward migrating Neptune.
We see from Table \ref{table:k1} that the 2:1 external resonance
has exceptionally large values of $\bar\epsilon$ compared
to the 3:2 and 4:3 resonances.  This is because $\delta_{1,0}$
is small due to the addition of the indirect term.

We first consider the critical migration rate allowing capture
and compare the one we predict here with that found from numerical
studies.
Migration rates are often given in terms of the time
it takes to cross the range of radius covered during the entire migration.
This is typically a few AU for
Neptune's migration \citep{ida00,chiang02}.  To compare
migration rates to the critical one estimated above, we must
first convert rates into our system of units. 
For $GM_\odot = 1$ 
and radii in units of Neptune's semi-major axis, $a_N$, we multiply
timescales by $\sqrt{a_N^3/G M_\odot} = 26.1$ years.
A migration rate of a few AU in $10^7$ years
corresponds to 
$\dot a_p \sim 2.6\times 10^{-7}
   \left({10^7 {\rm yr}\over t_{migrate}}\right)$.
Since $n_p \propto a_p^{-3/2}$ this corresponds to 
$\dot n_p \sim 3.9\times 10^{-7}
   \left({10^7 {\rm yr}\over t_{migrate}}\right)$.
The critical planet migration rate (listed in Table \ref{table:k1})
for Neptune's 2:1 resonance
is $\dot n_{p,crit} \approx 0.54 \mu^{4/3} \approx 1.0 \times 10^{-6}$.
We find that
\begin{equation}
{\dot n_p \over \dot n_{p,crit}}  \approx  0.4  
   \left({10^7 {\rm yr}\over t_{migrate}}\right).
\end{equation}
\citet{chiang02} found that 2:1 resonance captured
at the 50\% level for $t_{migrate} = 10^7$ years but was much
less efficient, capturing only 15\% of particles 
at faster migration rates of $t_{migrate} \sim 10^6$ years. 
The same rise in capture probability at $t_{migrate}\sim 10^7$ years
was seen by \citet{ida00}.
The sharp drop in capture probability 
is consistent with our predicted limit for the critical
migration rate.  
We find that we can account for the trends seen 
in the numerical studies of \citet{ida00,chiang02}
and confirm the theoretical explanation of  \citet{friedland01}.
We note that the transition from a probability of 50\% to 15\% occurs
over a fairly large range of drift rates.
In our toy model we could account for such a smooth transition
with initial particle eccentricity near $e_{lim}$.
However because the capture probability drops steeply for $e_0 > e_{lim}$
this explanation would require fine tuning of the initial particle
eccentricity distribution.
We note that
it is impossible to zero the eccentricity of a particle
in a simulation because of other perturbations. 
Also, because we have dropped most cosine terms in Equation (\ref{hamgen}), 
We have neglected these other perturbations in our Hamiltonian model.
Consequently it is difficult for us to compare the initial
eccentricity distribution of a simulation to the distribution
in our momentum $\bar \Gamma$.

We now consider the role of the corotation resonance.
From table \ref{table:k1} we find
$\bar \epsilon =  6.6 \mu^{-1/3} e_p$.
For Neptune $\mu =5.1 \times 10^{-5}$  
and we find that $\bar \epsilon \approx  180  e_p$.
For Neptune's current eccentricity $e_p \sim 0.008$ this places 
$\bar \epsilon \sim 1.4$.
This is somewhat above the critical corotation strength value 
allowing capture into the 2:1 resonance
according to figure \ref{fig:a1n} when the resonances are on top
of each other.
We need to consider the separation between the resonances; 
$\bar c = 1.18 \mu^{1/3} = 0.04$.  However this is the separation
only if Neptune's longitude of perihelion does not precess.
Neptune's precession frequency is largely 
due to the solar system's eighth eigenvector that dominates this 
planet's secular motions \citep{nobili,applegate}.
Neptune's precession rate due to other planets
is a few times larger than that it induces on objects in its 2:1 resonance.
Consequently $\bar c$ could be larger 
a factor of a few and either positive or
negative depending upon the secular motion of the planet
when the migration took place.
Smaller values of planet eccentricity would allow the 2:1 to capture
whereas larger values would tend to make it more difficult.
Negative values of $\bar c$ would allow
the resonance to be in the temporary capture 
regime shown in Figure \ref{fig:a2n}a
whereas positive values of $\bar c$ would allow capture at 
Neptune's current eccentricity, provided the migration was
slow (Figure \ref{fig:a2n}b).
It is interesting to find that Neptune's eccentricity
is very near the critical value that would make this resonance
fail to capture.  This large value of $\bar \epsilon$ could
contribute to the intermediate capture probability seen
in simulations and
the moderate range of drift rates where this intermediate capture
occurs.

\subsection{Capture into the 2:1 and 3:1 Resonances of
Multiple Extrasolar Planet systems}

Three extrasolar multi-planet systems have two planets
locked in the 2:1 resonance, GL876, HD 82943, and HD128311 \citep{marcy}.
In each case the outer planet is more massive than the inner one.
The masses of the outer planet are 1.9, 1.6 and 3.2 
2 $M_J$ (Jupiter masses), respectively.
We assume that an outer planet has migrated inward and captured
the interior and lower mass planet into the 2:1 resonance 
(e.g., as explored previously by \citealt{kley04,kley05,moorhead05}).
The coefficients for this
situation are listed in the appendix and in Table \ref{table:k1}.
For an internal 2:1 mean motion resonance $\dot n_{p,crit} = 22.7 \mu^{4/3}$.
We relate the critical mean motion drift rate  to a critical
semi-major axis drift rate (with a factor of 2/3) and restore the units.
To capture an internal planet into the 2:1 resonance
a planet must have a migration rate slower than
\begin{equation}
\dot a_p \lesssim 15 \mu^{4/3} \left({GM_* \over a_p}\right).
\end{equation}
Using the period of the planet's orbit, $P = 2\pi \sqrt{a_p^3 \over GM_*}$,
we can relate the critical migration rate to a timescale,
$\tau = a_p/\dot a_p$, finding
\begin{equation}
\tau_{2:1} \gtrsim 0.4 \mu^{-4/3} P
\end{equation}
for the 2:1 resonance
For a $2 M_J$ mass planet, we find a migration timescale of longer than
1600 orbital periods is required for the 2:1 resonance to capture.
This limit is consistent with the timescales adopted for
migration in the simulations by \citet{kley04}.
We can consider the eccentricity limit
(taking the value from Table \ref{table:k1})
$e_{lim} \sim 1.5 \mu^{1/3} = 0.2$.
This implies that the initial eccentricity of the inner planet
(as long as it was below 0.2)
would probably not limit the capture probability into the 2:1 resonance.

We now consider capture into the 3:1 resonance.  The 55 CnC system
has two planets locked in the 3:1 resonance with outer planet $0.2 M_J$
and inner planet with $0.8 M_J$.  Even though the inner planet
is more massive we consider capture into an internal resonance
because the 3:1 internal resonance is stronger than the external one.  
This is a result of the contributions from the indirect terms.
Using values given in the appendix in Table \ref{table:k2} for 
an internal 3:1  resonance we find  $\dot n_{p,crit} = 0.6 \mu^{2}$
for the $e^2$ resonance and $\dot n_{p,crit} = 41 \mu^{4/3} e_p^{4/3}$ 
for the $ee'$ resonance, in both cases
for low initial particle eccentricity.
This corresponds to a migration timescale
\begin{equation}
\tau_{3:1,e2} \gtrsim 15  \mu^{-2} P
\end{equation}
for the $e^2$ resonance and
\begin{equation}
\tau_{3:1,ee'} \gtrsim 0.2 \mu^{-4/3} e_p^{-4/3} P
\end{equation}
for the $ee'$ resonance.
However the migration rate is 
less restrictive for the $e^2$ resonance if the inner planet has a moderate
eccentricity.  The limiting eccentricity is $e_{lim} \sim 0.2 \mu^{1/2}$.
For $\mu = 0.002$, we find $e_{lim} = 0.003$.  Consequently
the inner planet is likely to have $e > e_{lim}$.
In this case the migration timescale must be modified by the factor
given in Equation (\ref{eqnnphalf}) and the limiting
migration timescale would be $\sim 10$ times smaller or 
\begin{equation}
\tau_{3:1,ee'} \gtrsim 0.02 \mu^{-4/3} e_p^{-4/3} P.
\end{equation}
For $\mu = 0.0002$ to capture we find the migration timescale
must be longer than a few times $10^7$ orbits for the $e^2$ resonance
and $20000 e_p^{-4/3}$ orbits for the $ee'$ resonance. For a moderate
planet eccentricity of 0.2 this corresponds to a timescale of 
$2\times 10^5$ orbits.  

We have found here that capture into the 2:1 resonance by
the multiple planet extra solar systems does not
require a slow migration rate but capture into the 3:1 resonance does.
For the 3:1 resonance faster migrations are allowed
for the $ee'$ resonance than the $e^2$ resonance.
The limiting timescale is decreased if the 
planet eccentricities are not low prior to resonance capture.

\section{Summary and Discussion}

In this paper we have explored the problem
of resonance capture for mean motion resonances 
at fast or non-adiabatic drift rates.
We first studied the first and second order time dependent 
Hamiltonian system with one resonant term.
We find that for sufficiently low
initial particle momentum (or eccentricity), the transition between resonance 
capture and no capture is sharp, occurring over a narrow
range in drift rate.  
We give an expression (Equation \ref{eqnnpcrit})
which makes it possible to predict the critical planetary migration
rate (above which there is no capture) for first and
second order mean motion resonances in the general restricted
three body problem in the limit of low initial particle eccentricity.  
Expressions are given in the appendix for coefficients which
allow one to estimate the critical drift rate for any first or 
second order mean motion resonance.  Coefficients are evaluated
for strong resonances and listed in Tables \ref{table:k1} and \ref{table:k2}.
This generalizes upon previous analytical work by \citet{friedland01}
and provides a theoretical explanation for critical drift rates
measured numerically and their dependence on 
planet mass (e.g., \citealt{ida00,wyatt03}).

We have numerically measured the probability of capture
as a function of initial particle eccentricity.
We find that the transition between resonance 
capture and no capture is smoother, occurring over a larger
range in drift rate, for initial particle eccentricity of order
the limiting value ensuring capture in the adiabatic limit, $e_{lim}$.
The drift rate at which the capture probability is half
is not strongly dependent
on the initial particle eccentricity for first order resonances, and
the probability of capture drops rapidly 
for initial particle eccentricities exceeding
the limiting value, $e_{lim}$. For second
order resonances, we find that the 
drift rate at which the capture probability is half 
is higher when the initial particle eccentricity is higher.
Equation (\ref{eqnnphalf}) can be used to estimate
the half probability drift rate 
for initial eccentricities below $10 e_{lim}$.
At $e_0 \gtrsim 30  e_{lim}$ 
the capture probability drops below 1/2 at all drift rates.

In the limit of low initial particle eccentricity, 
we have considered the case of resonances containing
multiple subterms.
A first order resonance fails to capture when
the corotation resonance has unitless strength $\bar \epsilon \gtrsim 1$.
As this coefficient depends on planet eccentricity,
migrating, eccentric, low mass planets could have first order resonances
that fail to capture particles for this reason. 
A regime of intermediate capture probability also exists at high
drift rates and large corotation perturbation strength.
We have found that the resonance
separation, and order of encounter, affects the capture probability,
primarily when the corotation resonance is strong.
This implies that the capture probability is dependent upon
the precession rates of the longitude of periapse of 
both particle and planet.

Second order resonances contain three subterms.  As was
true for the first order resonances, the corotation
resonance fails to capture particles but can prevent
the other resonant terms from capturing particles if 
scale free parameters $\bar \epsilon_\xi \gtrsim 1$ or 
$\bar \epsilon_\chi \gtrsim 1$.  This implies
that above a certain planet eccentricity, second order resonances
fail to capture particles. 
Below this planet eccentricity the $e^2$-  and $ee'$-resonances
can capture particles.  When our coefficient $\bar \xi \lesssim 10^{-2} $
the $e^2$-resonance will capture particles (providing the drift rate is
sufficiently slow) and the capture behavior is
second order. For $\bar \chi < 0.1$ the $ee'$-resonance will capture particles
and the behavior is first order.  For nonzero planet eccentricity,
the $ee'$ subresonance (which behaves like a first
order resonance) may more easily capture particles at
faster drift rates than the $e^2$ resonance.
For $\bar \chi$ or $\bar \xi$  of order 1, a regime of 
intermediate capture probability exists at high drift rates.
For second order resonances the subresonance
separation, and order of encounter also affects the capture probability.

A number of effects have been proposed to account for
reduction in capture probabilities compared to those
predicted via adiabatic theory,
e.g., \citet{zhou02} showed that stochastic or jumpy
migration would allow particles to escape resonances.
Here we have shown that rich dynamics in
the non-adiabatic limit allows
particles to escape resonance capture.
We have shown that corotation terms can reduce the capture
probability.  For second order resonances, resonant subterms
can interfere, again producing a regime of intermediate capture
probability.
For first order resonances, the half probability
drift rates are not strongly dependent on the initial
particle eccentricity, and the probability of capture drops
rapidly above a limiting initial eccentricity.
However for second order resonances the half probability 
drift rate is higher for initial particle eccentricity
near the limiting value.  Consequently we expect that second
order resonances should have larger regimes 
of intermediate capture probability 
in range of drift rate and initial particle eccentricity.

We have applied our understanding to the problem of capturing
twotinos via Neptune's migration.
We find that  the eccentricity of Neptune is sufficiently
high that the 2:1 resonance could fail to capture particles.
Certainly if Neptune's eccentricity were any higher during
migration its 2:1 resonance would not have captured particles
efficiently. It is interesting to find that Neptune's eccentricity
is very near the critical value that would make this resonance
fail to capture particles.

We have applied our framework toward predicting mininum migration
timescales allowing extra solar multiple planet systems to capture
into the 2:1 or 3:1 resonances.  We find that a migration timescale
of greater than a few thousand orbital periods is required to allow
capture into the 2:1 resonance for three systems.
However a much longer timescale,
$\sim 10^7$, orbital periods is required to allow capture into
the 3:1 resonance for the 55 Cnc planetary system.  
The migration timescale can be somewhat reduced if
the planets are on moderately eccentric orbits subsequent to 
migration.


In this work we have extended the theory of resonant
capture for drifting Hamiltonian systems 
to the non-adiabatic limit
and to systems with multiple resonant subterms. 
We have provided a theoretical framework to predict resonance
capture probabilities.  However this framework is based on 
numerical integration of a simplistic two-dimensional Hamiltonian 
model and so may not accurately
represent the full complex dynamical systems.
Direct numerical integration of these systems must be carried out to 
test the validity and accuracy of the expressions given in this paper.
The exploration done here could also in future be extended via numerical
study of modified quasi Hamiltonian toy models (e.g., \citet{gomes97}) 
to better cover systems with drift induced by 
non-conservative forces such as gas drag or Poynting-Robertson
drag.
This work could also be extended to cover motions out of the plane
and high eccentricity systems.

\acknowledgments

I thank the Research School of Astronomy and Astrophysics of the 
Australian National University 
and Mount Stromlo Observatory 
for hospitality and support during Spring 2005.
Support for this work was in part
provided by National Science Foundation grant AST-0406823,
and the National Aeronautics and Space Administration
under Grant No.~NNG04GM12G issued through 
the Origins of Solar Systems Program.  
Support was also provided by
the National Science Foundation to the Kavli Institute
for Theoretical Physics under Grant No.~PHY99-07949.

\appendix 

\section{Coefficients for Internal and External resonances}

\subsection{External resonances}

For external first order resonances, 
\begin{eqnarray}
K(\Lambda, \psi; \Gamma,\gamma) 
   &  = & 
   a \Lambda^2 + b \Lambda + c \Gamma 
  \qquad  
\label{hamk1}
\\
& &  + \delta_{1,0} \Gamma^{1/2} \cos{(\psi - \varpi)}
   + \delta_{1,1} \cos{(\psi - \varpi_p)} \nonumber
\end{eqnarray}
with
\begin{eqnarray}
\delta_{1,0}   &=& -\mu \sqrt{2} \alpha^{5/4} f_{31}    \nonumber \\
\delta_{1,1}   &=& -\mu e_p \alpha       f_{27}.
\label{delta1}
\end{eqnarray}
Coefficients $a,b,$ and $c$ are given in 
Equations (\ref{eqnab}) and (\ref{eqnc}).
The $f_i$ are functions of the Laplace coefficients
and are evaluated at $\alpha$ with index $j$ 
using expressions from the appendix by \citet{M+D}. 

The above expressions only include direct terms.
For the 2:1 resonance the indirect term contributes and
\begin{equation}
\delta_{1,0}(2:1) =  
  -\mu \sqrt{2} \alpha^{1/4} 
   \left( 
          \alpha f_{31} 
        - {1 \over 2 \alpha} 
   \right)
\label{delta2to1}
\end{equation}
The near cancellation of the direct
and indirect terms makes second order terms 
important for the 2:1 resonance \citep{friedland01, murray_clay05}.
It may be useful to recall
the maximum initial particle momentum 
or initial eccentricity, $e_{lim}$ ensuring capture in the adiabatic
limit. This corresponds to 
\begin{equation}
e_{lim} = \sqrt{2 \bar \Gamma_{0,lim}} \alpha^{1/4} 
    \left|{ \delta_{1,0} \over a} \right|^{1/3}
\label{eqnecrit}
\end{equation}
where $\bar \Gamma_{0,lim} = 3/2$.

For external second order resonances or $k=2$,
\begin{eqnarray}
K(\Lambda, \psi; \Gamma,\gamma) 
   &  = & 
   a \Lambda^2 + b \Lambda + c \Gamma 
  \qquad  
\\
& &  + \delta_{2,0} \Gamma       \cos{(\psi - 2 \varpi)}\nonumber \\
& &  + \delta_{2,1} \Gamma^{1/2} \cos{(\psi - \varpi - \varpi_p)} \nonumber \\
& &  + \delta_{2,2} \cos{(\psi - 2 \varpi_p)} \nonumber
\end{eqnarray}
where
\begin{eqnarray}
\delta_{2,0}  &=& -\mu 2   \alpha^{3/2} f_{53}    \nonumber \\
\delta_{2,1}  &=& -\mu e_p \sqrt{2} \alpha^{5/4}  f_{49}   \nonumber \\
\delta_{2,2}  &=& -\mu e_p^2 \alpha          f_{45}.
\end{eqnarray}
Coefficients $a,b$ and $c$ are given in 
Equations (\ref{eqnab}) and (\ref{eqnc}).
For the 3:1 resonance the indirect term contributes and 
\begin{equation}
\delta_{2,0} (3:1) = -\mu 2 \alpha^{1/2}
           \left({ \alpha f_{53}   
                  -{3  \over 8 \alpha}}\right).
\end{equation}
For second order resonances capturing into the $e^2$ subterm
\begin{equation}
e_{lim,\xi} = \sqrt{2 \bar \Gamma_{0,lim}} \alpha^{1/4} 
    \left|{ \delta_{2,0} \over a} \right|^{1/2}
\end{equation}
where the critical scale free momentum $\bar \Gamma_{0,lim} = 1/8$.
For those capturing into the $ee'$ subterm
\begin{equation}
e_{lim,\chi} = \sqrt{2 \bar \Gamma_{0,lim}} \alpha^{1/4} 
    \left|{ \delta_{2,1} \over a} \right|^{1/3}
\end{equation}
where the critical scale free momentum $\bar \Gamma_{0,lim} = 3/2$.

\subsection{Internal resonances}

To make our theory appropriate for internal resonances (external
perturber) we
would consider $j:j+k$ resonances and change the coefficients to 
\begin{eqnarray}
a &=& -{3 \over 2} j^2 \alpha^{-2}      \nonumber \\
b &=& -(j +k) (n_p - 1) \nonumber \\
c &=& -{\mu 2 f_2} \alpha^{-1/2} \nonumber \\
\delta_{1,0}  &=& -\mu \sqrt{2} \alpha^{-1/4} f_{27}    \nonumber \\
\delta_{1,1}  &=& -\mu e_p f_{31}  \nonumber \\
\delta_{2,0}  &=& -\mu 2   \alpha^{-1/2} f_{45}    \nonumber \\
\delta_{2,1}  &=& -\mu e_p \sqrt{2} \alpha^{-1/4}  f_{49}   \nonumber \\
\delta_{2,2}  &=& -\mu e_p^2       f_{53}
\label{eqn_internal}
\end{eqnarray}
where $\alpha \equiv a/a_p$ and
we have used the approximation $e^2 \sim 2 \Gamma/L 
\sim 2 \Gamma \alpha^{-1/2}$.
The $c$ term describes secular precession of the longitude of periapse
and depends on the function $f_2$ given in 
the appendix by \citet{M+D} and is evaluated at $\alpha$ with index $j=0$.
For internal resonances the other $f_i$ functions 
are evaluated at $\alpha$ with index $j+k$ 
using expressions from the appendix by \citet{M+D}. 
For the 2:1 and 3:1 resonances the indirect terms contribute and 
\begin{equation}
\delta_{1,1} (2:1) =  
  -\mu e_p \left( f_{31} - 2 \alpha \right)
\end{equation}
\begin{equation}
\delta_{2,2} (3:1) =  -\mu e_p^2 
           \left( f_{53} - {27 \over 8} \alpha \right).
\end{equation}
The maximum initial particle eccentricity ensuring capture in the 
adiabatic limit for first order resonances
\begin{equation}
e_{lim} = \sqrt{2 \bar \Gamma_{0,lim}} \alpha^{-1/4} 
    \left|{ \delta_{1,0} \over a} \right|^{1/3}
\end{equation}
where $\bar \Gamma_{0,lim} = 3/2$.
For second order resonances capturing into the $e^2$ subterm
\begin{equation}
e_{lim,\xi} = \sqrt{2 \bar \Gamma_{0,lim}} \alpha^{-1/4} 
    \left|{ \delta_{2,0} \over a} \right|^{1/2}
\end{equation}
where the critical scale free momentum $\bar \Gamma_{0,lim} = 1/8$.
For those capturing into the $ee'$ subterm
\begin{equation}
e_{lim,\chi} = \sqrt{2 \bar \Gamma_{0,lim}} \alpha^{-1/4} 
    \left|{ \delta_{2,1} \over a} \right|^{1/3}
\end{equation}
where the critical scale free momentum $\bar \Gamma_{0,lim} = 3/2$.
The drift rate for a capture probability
of 1/2 given in Equation (\ref{eqnnphalf})
must be modified for second order internal resonances;
\begin{equation}
|\dot n_{p,1/2}| \sim  0.5(j-2) \delta_{2,0}^{2} 
      \left( 1 + {\alpha^{1/2} e_0^2 a\over 2.4 \times 10^{-4} \delta_{2,0} } 
        \right)^{0.25}.
\end{equation}
This expression is valid for initial particle eccentricity
$e_0 \lesssim 10 e_{lim,\xi}$.

The coefficients for strong internal and external resonances
are listed in Tables \ref{table:k1} and \ref{table:k2}.

\section{Migration of dust via Poynting-Robertson Drag}

In this paper we have considered an varying Hamiltonian
system.  However there may be some similarities between
this system and the slowly drifting dissipative systems.
We add relations that allow the reader to
approximately predict the critical drift rates for dust
spiraling inward under Poynting-Robertson drag.
In the case of Poynting Robertson drag dust particles in a circular
orbit decay on a timescale proportional $\beta^{-1}$  where $\beta$ is the
ratio of radiation to gravitational (from the star) forces.
It is convenient to write
\begin{equation}
\beta \sim {0.2 \over s_{\mu m}} 
  \left({L_* \over L_\odot}\right)
  \left({M_* \over M_\odot}\right)^{-1}
\end{equation}
where
$s_{\mu m}$ is the radius of the particle in $\mu m$
and $L_*$ is the luminosity of the star \citep{sicardy93}.
The drag force leads to a slow increase in the mean motion
\begin{equation}
\dot n \sim { 3 \alpha^{1/2} \beta  \over  c_l} 
\end{equation}
where $c_l$ is the speed of
light in units of the planet's velocity
or divided by $\sqrt{GM_*/a_p}$.
The value of our coefficient $b$ is not important, as long
as it passes through zero on resonance.  However, its drift
rate or $\dot b$ is important.
At resonance $j n = (j-k) n_p$ and we can relate
the drift rate of the particle spiraling inward
to a system of a planet migrating outward
considered in the previous sections.
We replace $\dot n_p$ with $\dot n$, finding an effective 
drift coefficient 
\begin{equation}
\dot b =  {3 j  \alpha^{1/2} \beta \over c_l}.
\end{equation}
The rescaled speed of light
\begin{equation}
c_l \approx 10^{4}    \left( {M_* \over M_\odot} \right)^{-1/2}
                \left( {a_p \over 1AU    } \right)^{ 1/2}.
\end{equation}
Consequently we can write
\begin{equation}
\dot b =  0.6 \times 10^{-4}  j \alpha^{1/2}   
      s_{\mu m}^{-1} 
     \left({L_* \over L_\odot}\right)
     \left( {M_* \over M_\odot} \right)^{-1/2}
     \left( {a_p \over 1AU    } \right)^{-1/2}.
\end{equation}
The above relation can be used to approximately determine the minimum
size particles that can be captured into resonances using the formulation
presented in previous sections of this paper.

{}
\vskip 3 truein

\renewcommand{\thefigure}{\arabic{figure}}
\setcounter{figure}{0}

\renewcommand{\thetable}{\arabic{table}}
\setcounter{table}{0}

\begin{figure*}
\plottwo{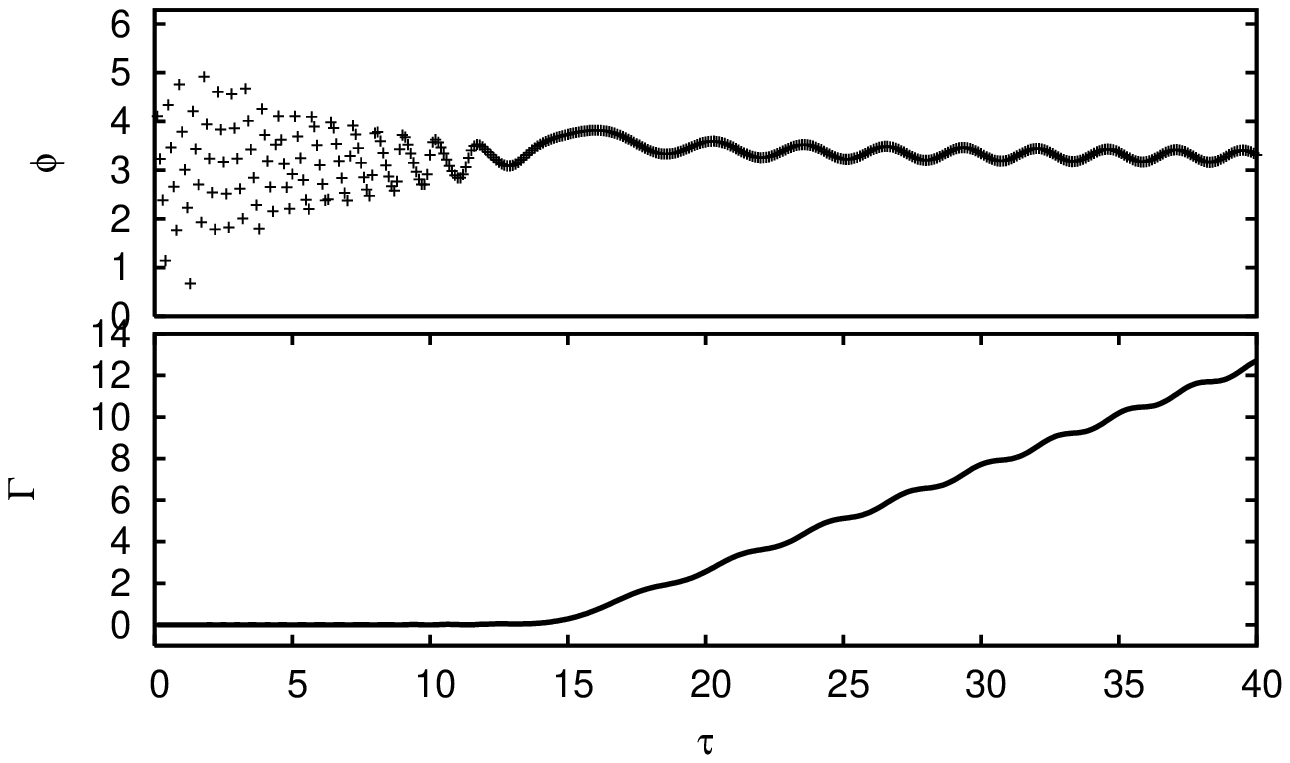}{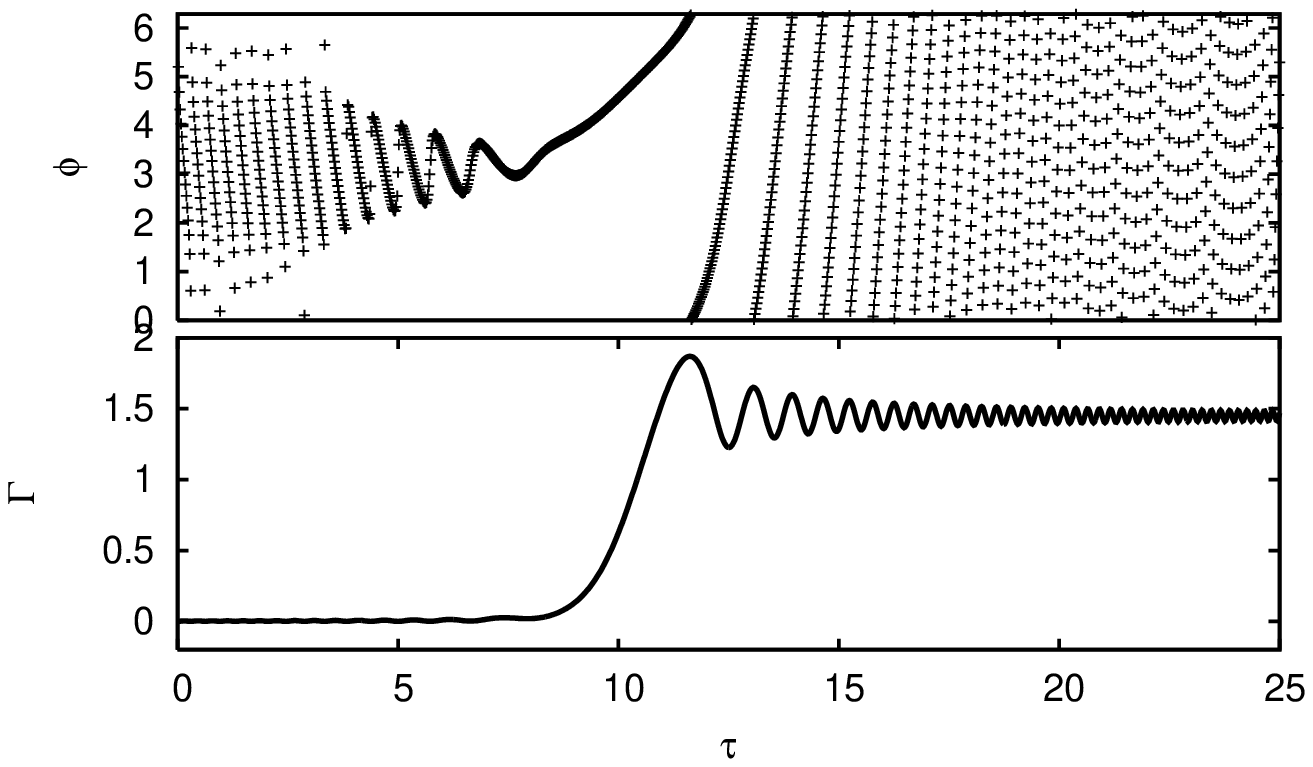}
\vskip 0.15truein
\figcaption{
a) This integration shows a resonant capture.
We show different behavior for the drifting system with
one resonant term,
Equation (\ref{eqnKp}), as it passes through resonance.  
For these integrations the resonance is first order ($k=1$),  
and the initial momentum $\bar \Gamma(t_0) = 10^{-4}$.  
The solid lines shows $\bar \Gamma$.
The dots show the resonant angle $\phi$.
After capture at a time $\tau \sim 15$, the resonant angle
librates about $\pi$ and the momentum $\bar \Gamma$ slowly increases.
The rate of increase is set by the drift rate; for this integration
is ${d\bar b\over d\tau}=1$.
b) No capture takes place in this integration which has
a higher drift rate of ${d\bar b\over d\tau}=2.3$.
The resonant angle $\phi$ circulates during  the entire integration.
The momentum oscillates about a fixed value before and after
resonance.  There is an increase or jump in the mean momentum 
as the system passes through resonance at $\tau \sim 10$.
\label{fig:bfig}
}
\end{figure*}

\begin{figure*}
\plotone{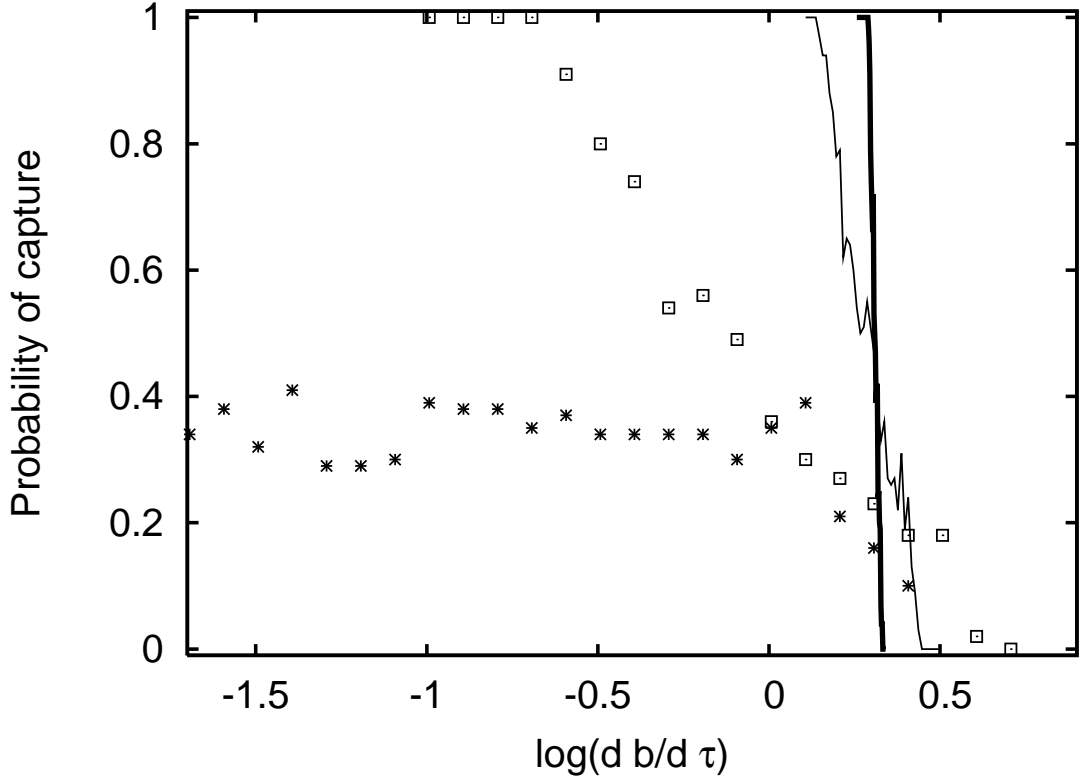}
\figcaption{
Capture probabilities for a Hamiltonian system Equation (\ref{eqnKp}) 
with one first order ($k=1$) resonant term as a function of drift rate 
$\left|{d \bar b \over d \tau}\right|$ and initial momentum $\bar \Gamma (t_0)$.
Note that low $\bar \Gamma(t_0)$ corresponds to low initial eccentricity.
The thick and thin solid lines shows the capture probability for 
$\bar \Gamma (t_0) = 10^{-4}$ and $10^{-1}$ respectively.  
The squares and stars show the capture probability for 
$\bar \Gamma (t_0) =  1 $ and $2.3$ respectively.  
In the adiabatic limit for
$\bar \Gamma(t_0)<1.5$ the capture probability is 1.
For $\bar \Gamma(t_0) = 2.3$ the capture probability is intermediate
for low drift rates and approaches a constant
value as the system becomes more adiabatic.
For initial momentum low ($10^{-4}$) the transition between 100\% capture and
0\% capture is extremely sharp.   We find that if
the initial momentum is $\sim 1$ then there is a regime 
or a range of drift
rates where the capture probability is intermediate.
With a change of scale, all first order resonances can be put
in the form of Equation (\ref{eqnKp}). Consequently 
the probabilities shown here can be used to estimate
the capture probability in the non-adiabatic limit for
any migrating first order resonance.
\label{fig:simplek1}
}
\end{figure*}

\begin{figure*}
\plotone{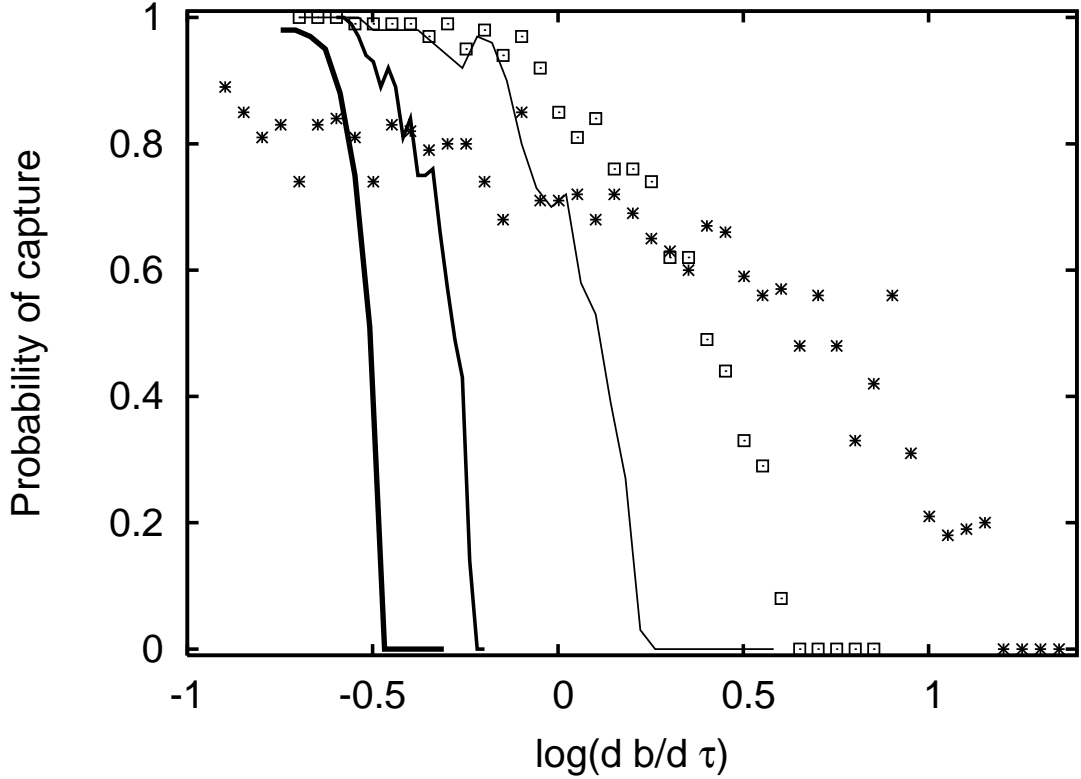}
\figcaption{
Similar to Figure \ref{fig:simplek2} except for the second
order resonances (Equation \ref{eqnKp} with $k=2$).
The thick, intermediate and thin solid lines show the capture probability for 
$\bar \Gamma (t_0) = 10^{-6},10^{-4}$ and $10^{-2}$ respectively.  
The squares and stars show the capture probability for 
$\bar \Gamma (t_0) = 0.1 $ and $1.0$ respectively.  
Capture is ensured in the adiabatic limit for
$\bar \Gamma(t_0) < 1/8$.
As was true for the first order resonances
the transition between capture and no capture is steeper (covering
a narrower range in drift rate) for low initial momentum.
For second order resonances 
the drift rate for a capture probability of 0.5
depends on the initial momentum.
\label{fig:simplek2}
}
\end{figure*}

\begin{figure*}
\plottwo{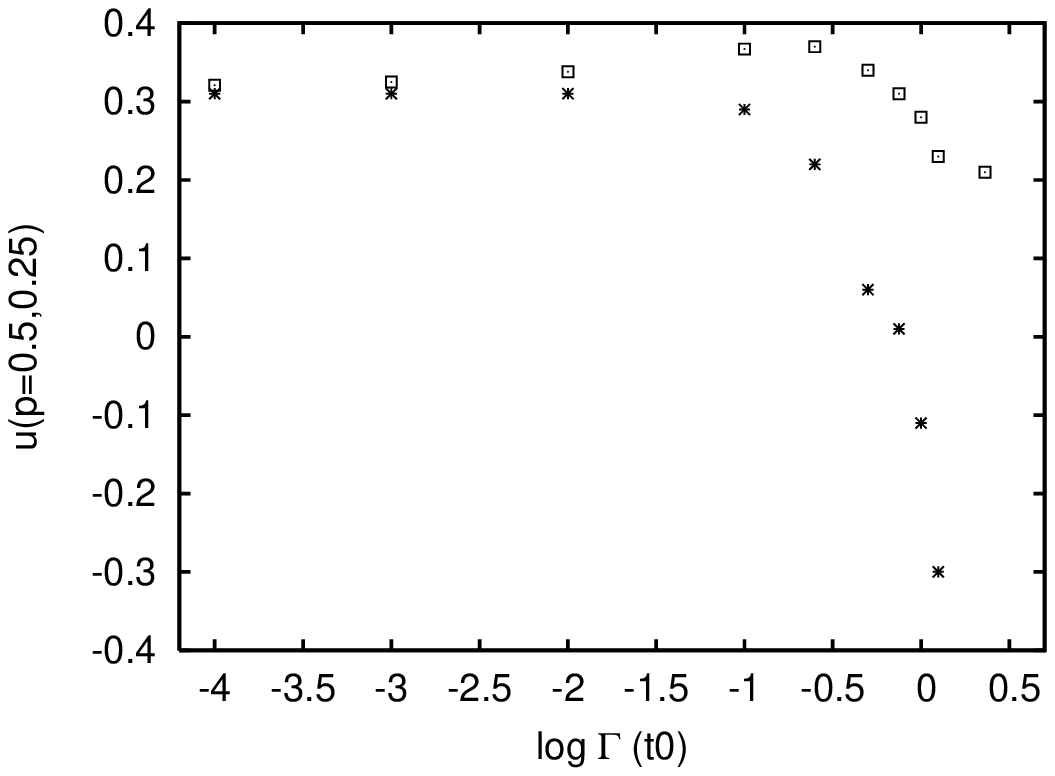}{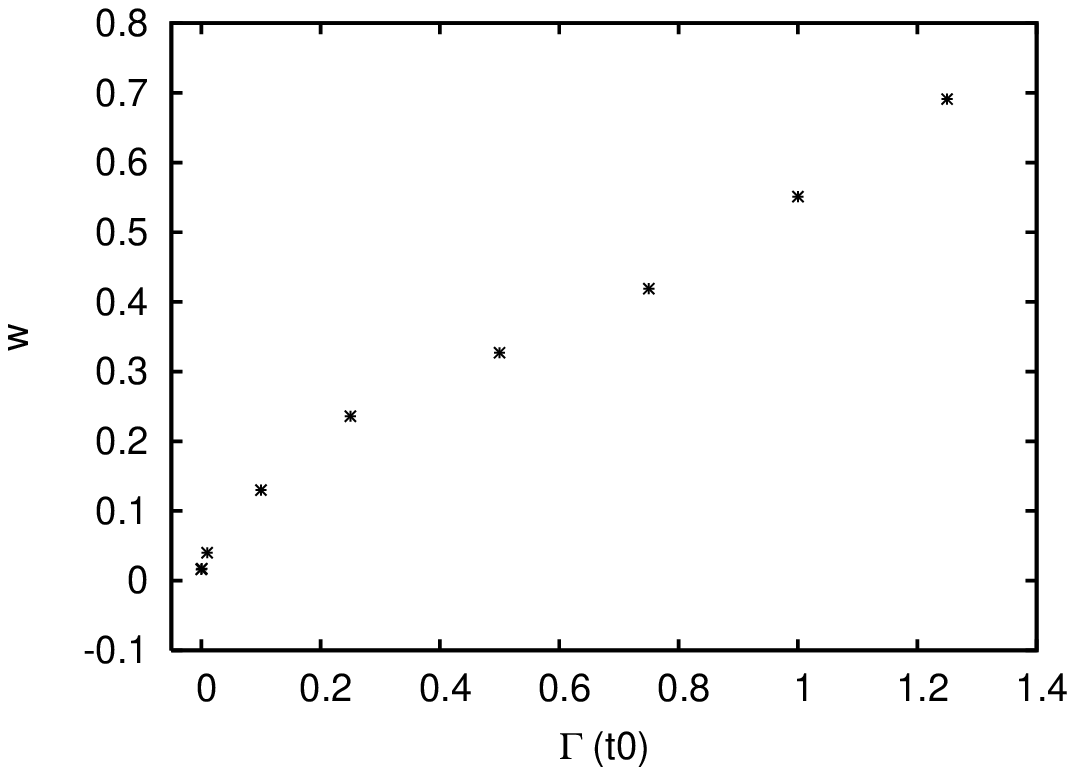}
\figcaption{
a) The drift rate at which the probability of capture is 1/2 (stars) 
and 1/4 (squares)
as a function of initial momentum ($\bar \Gamma (t_0)$ )
for first order resonances ($k=1$).
The $x$-axis shows $\log_{10} \bar \Gamma(t_0)$.
The $y$-axis shows $\log_{10}\left|{d \bar b \over d \tau}\right|$.
b) Width of the drop in probability  as a function of initial momentum.
\label{fig:gamk1}
}
\end{figure*}

\begin{figure*}
\plottwo{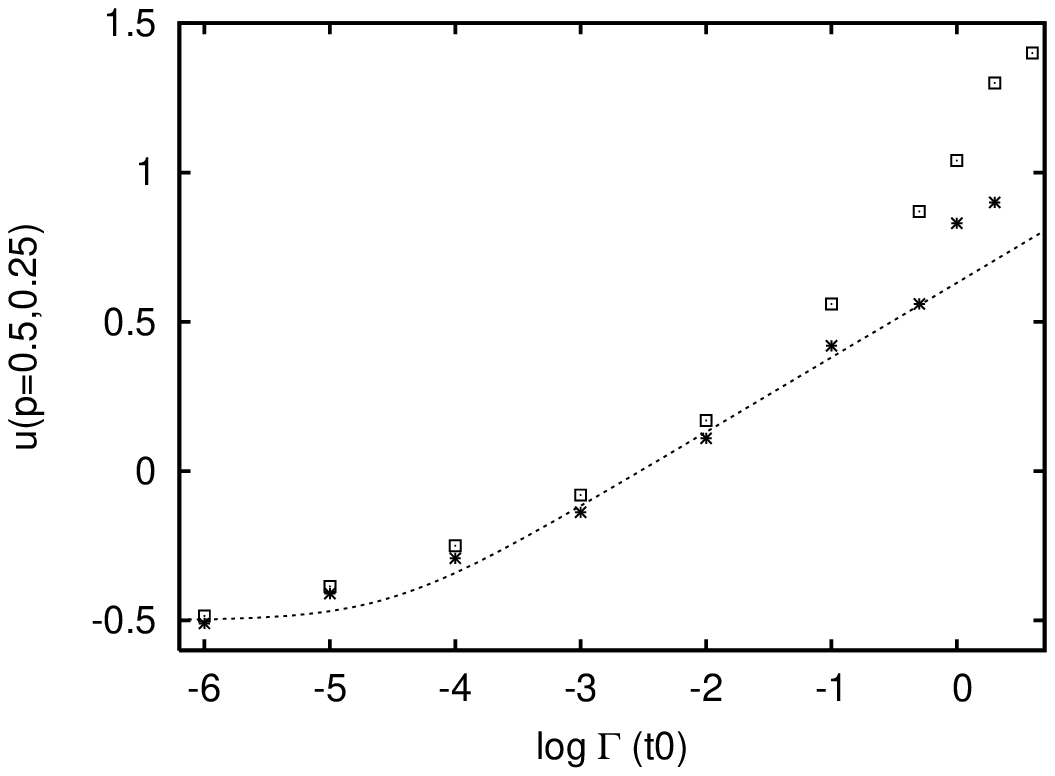}{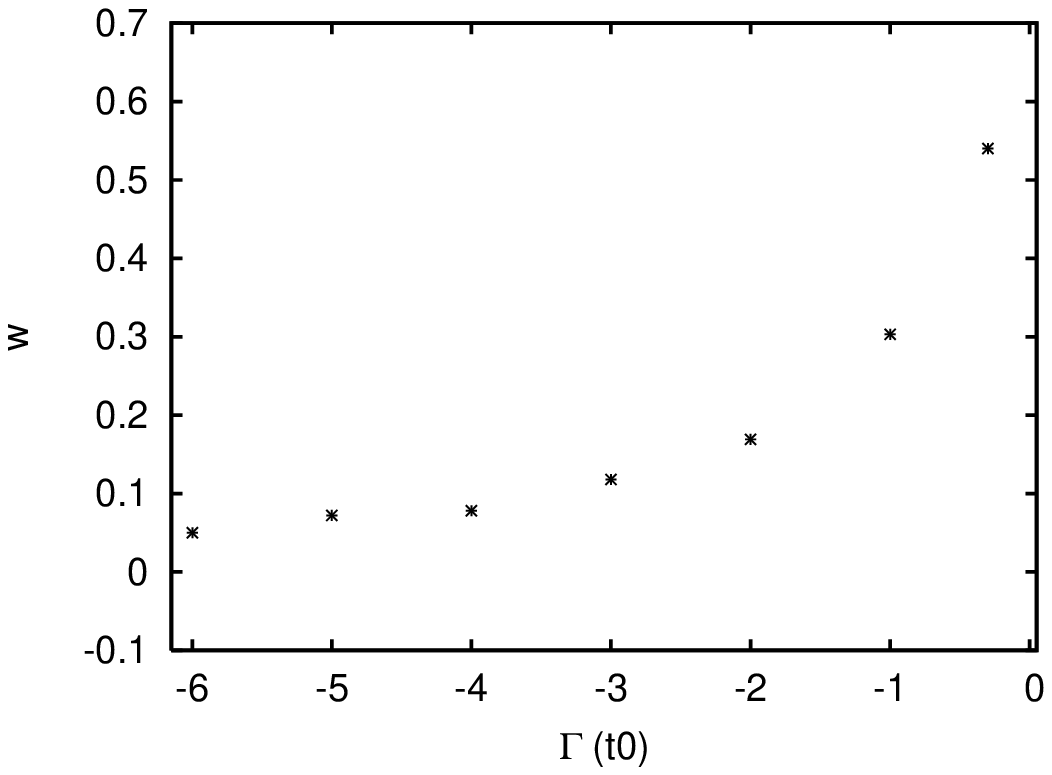}
\figcaption{
a) The drift rate at which the probability of capture is 1/2 (stars) 
and 1/4 (squares)
as a function of initial momentum for first second order
resonances ($k=2$).
The $x$-axis shows $\log_{10} \bar \Gamma(t_0)$.
The $y$-axis shows $\log_{10}\left|{d \bar b \over d \tau}\right|$.
The dotted line shows the function given in Equation (\ref{eqnhalf}).
b) Width of the drop in probability  as a function of initial momentum.
\label{fig:gamk2}
}
\end{figure*}

\begin{figure*}
\epsscale{0.50}
\plotone{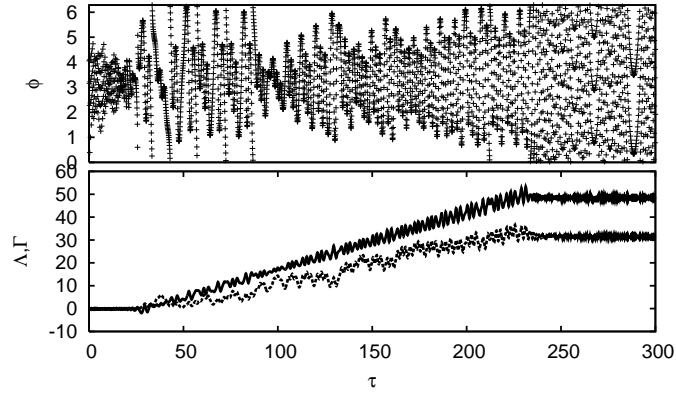}
\vskip 0.15truein
\figcaption{
The Hamiltonian system with two resonant terms can also
exhibit temporary captures.
Here Equation (\ref{eqnKp}) is integrated as it passes through resonance.  
with $\bar \epsilon = 1.8$ , ${d \bar b \over d \tau} = 0.5$, 
$\bar c =  0.9$ and $ \bar \Gamma (t_0) = 10^{-4}$.
The momenta $\bar \Lambda$ and $\bar \Gamma$ are shown as a solid and dotted
lines, respectively.
\label{fig:partcap}
}
\end{figure*}

\begin{figure*}
\plotone{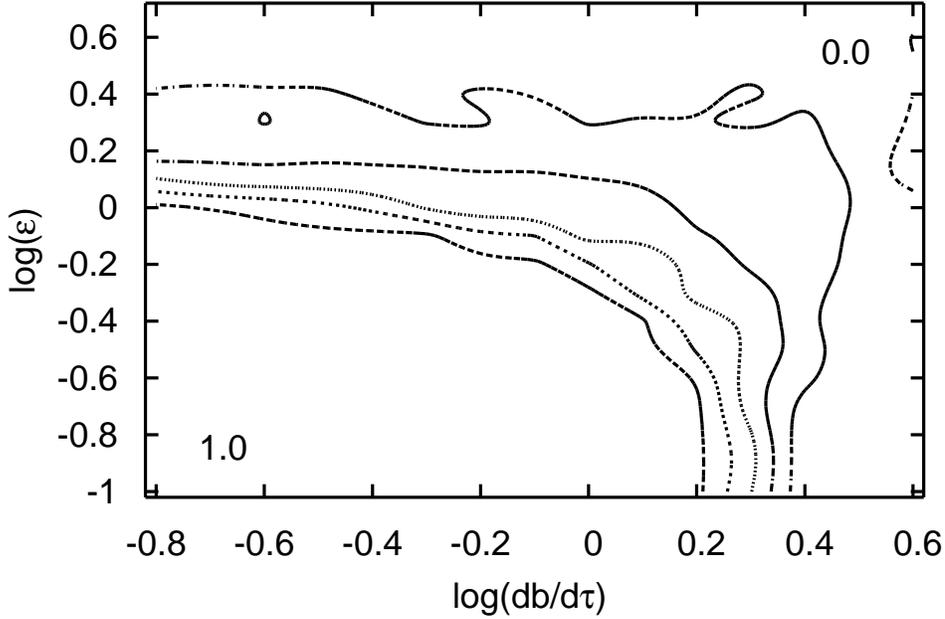}
\figcaption{
The capture probability for a first order resonance ($k=1$) is
shown as a function of drift rate $|{d \bar b\over d\tau}|$
and strength of secondary resonance $\bar \epsilon$.
Contours are shown at probabilities of $0.15, 0.35, 0.55, 0.75$ and $0.95$.
The numbers 0.0 and 1.0 on the top right and lower left, respectively
are placed to make it clear where the probability of capture is near zero
and near 1.
Equation (\ref{eqnvarne}) was numerically integrated with resonance
separation $\bar c=0$,
and low initial momentum $\bar \Gamma (t_0) = 10^{-4}$.
Each probability was measured from 100 different trials with
randomly chosen initial angles.
The $x$-axis shows $\log_{10}|{d \bar b\over d\tau}|$ and the $y$-axis
shows $\log_{10}({\bar \epsilon})$.
For low values of corotation perturbation strength,
$\bar \epsilon$, the transition between capture
and no capture happens at the critical value of ${d \bar b \over d \tau}$ 
estimated for the case of a single perturbation.
Near the critical drift rate the additional resonant 
perturbation 
can cause a moderately large region with an
intermediate probability of capture.
At lower drift rates, the corotation resonance prevents capture
for $\bar \epsilon \gtrsim 1$.
To the upper right on this plot (large drift rates
and large $\bar \epsilon$ values), temporary capture
can take place, making it more difficult to measure
absolute capture probabilities.
\label{fig:a1n}
}
\end{figure*}

\begin{figure*}
\plottwo{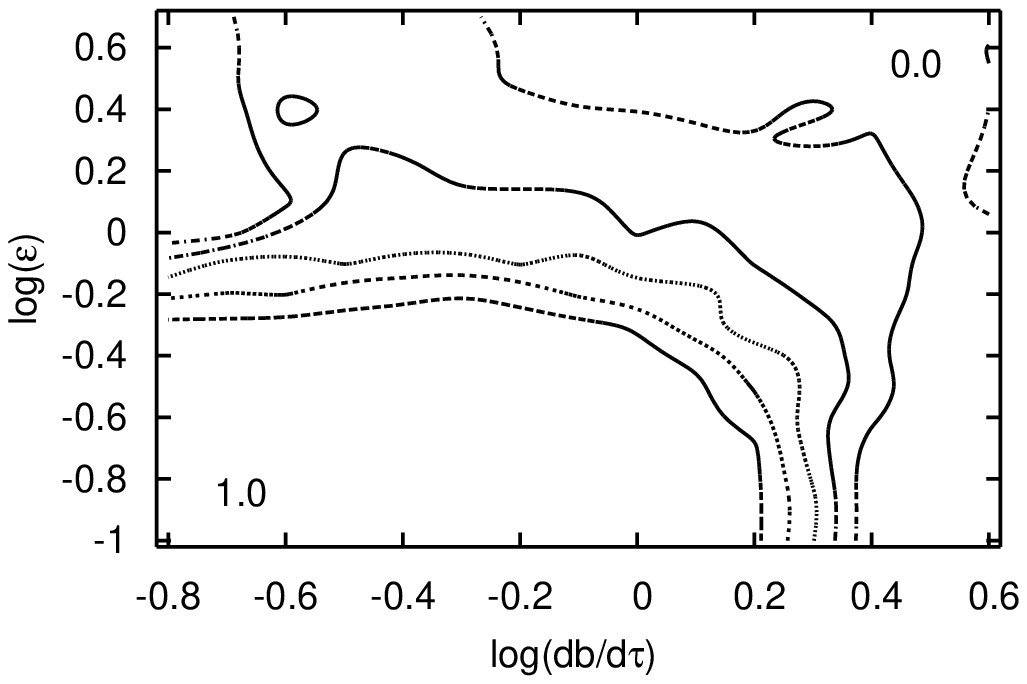}{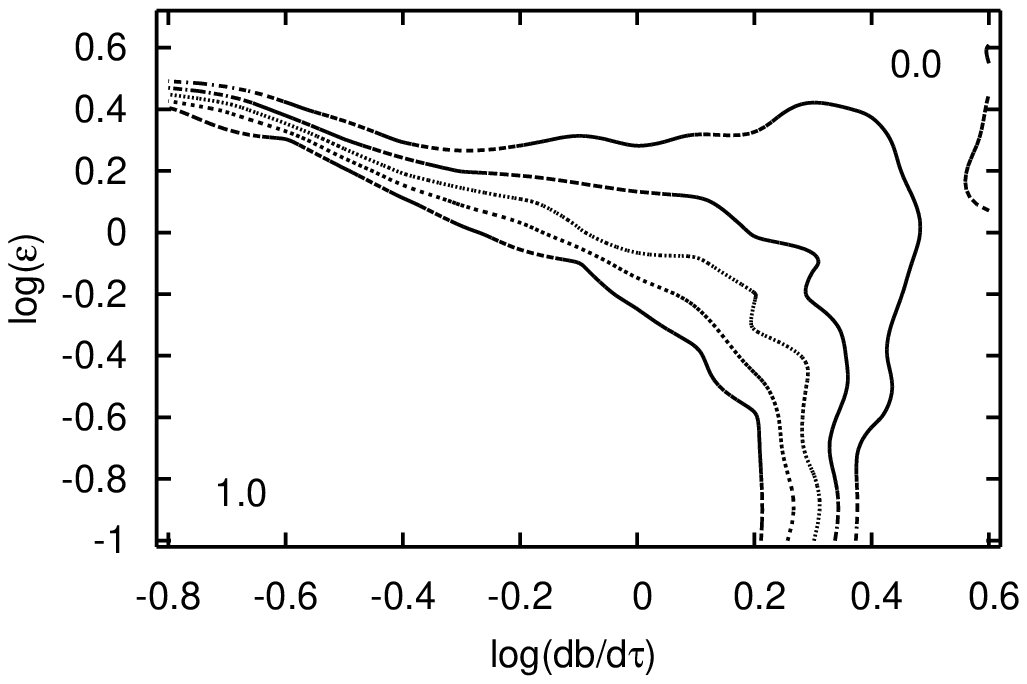}
\figcaption{
a) Same as Fig \ref{fig:a1n} except the resonance
separation $\bar c = -0.1$.
The frequency of the corotation term 
is shifted so that this resonance is encountered after
the $e$-resonance. 
The onset of the corotation resonance can kick the particle 
out of resonance.
b) Same as Fig \ref{fig:a1n} except the resonance
separation $\bar c = 0.1$.
The frequency of the corotation term 
is shifted so that this resonance is encountered first.
At lower drift rates larger $\bar \epsilon$ is required
to prevent captures.
These figures show that the subresonance separation
can influence the capture probability. 
The subresonance separation is set by the difference
between the planet and particle's precession rate of longitude
of periapse.
\label{fig:a2n}
}
\end{figure*}

\begin{figure*}
\plottwo{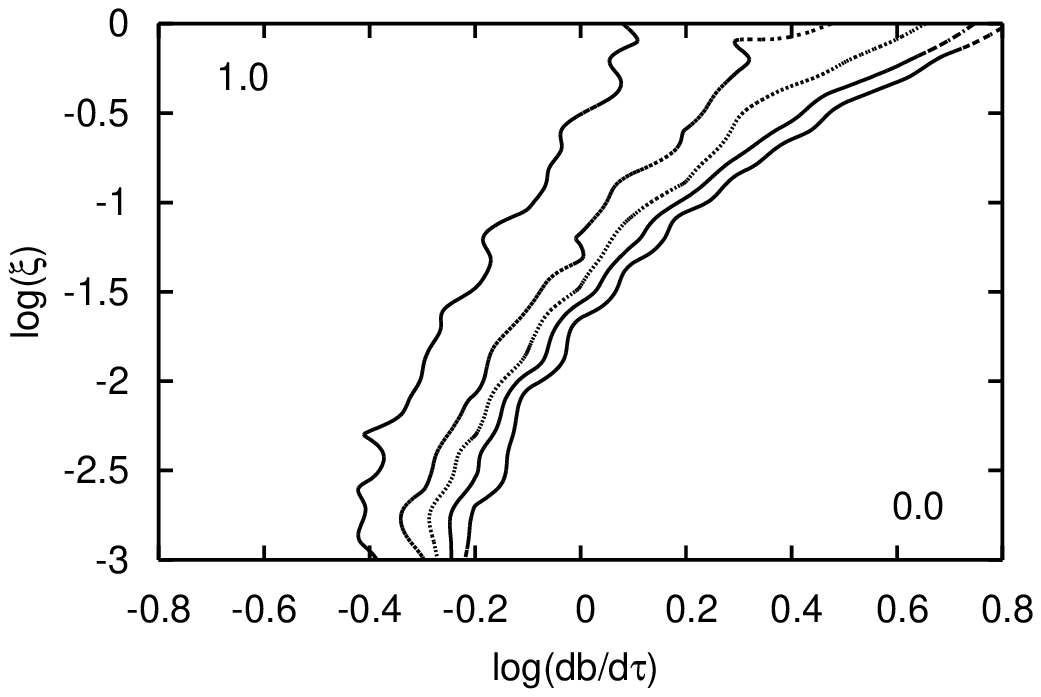}{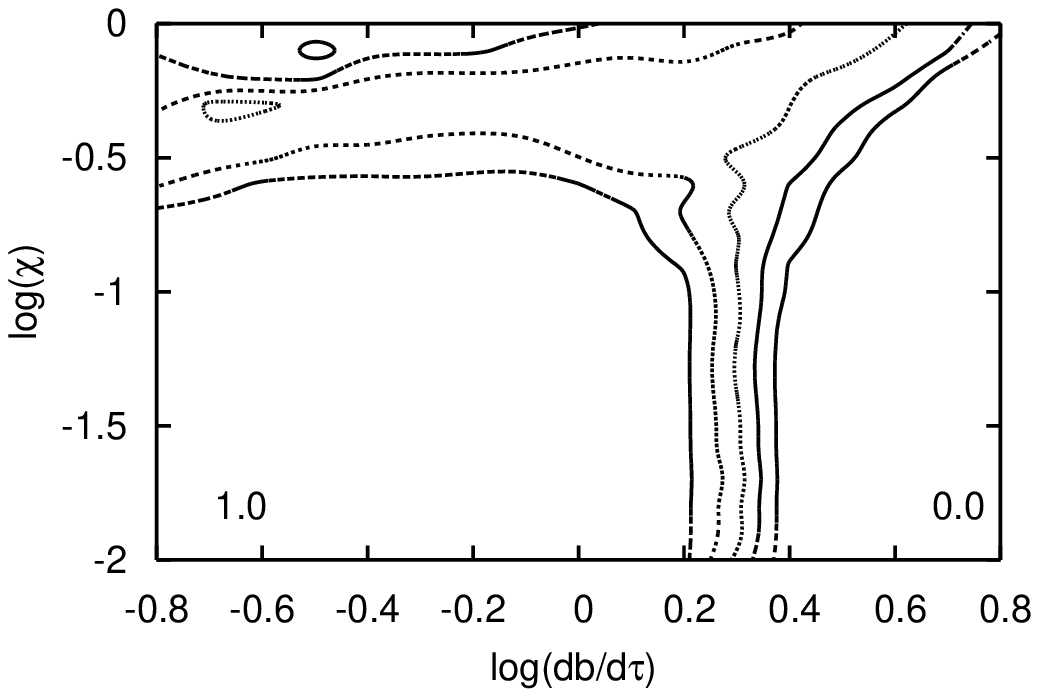}
\figcaption{
a) The capture probability for a second order resonance ($k=2$)
as a function of drift rate and $ee'$-resonance strength.
This figure is similar to Fig. \ref{fig:a1n}.
The Equation (\ref{eqnvarnexi}) was numerically integrated with 
resonance separation $\bar c_\xi=0$,
no corotation term, $\bar\epsilon_\xi = 0$,
and initial momentum $\bar \Gamma (t_0) = 10^{-6}$.
Drift rate and $\bar \xi$ were varied.
The $x$-axis is $\log_{10}|{d \bar b_\xi \over d\tau}|$
and the $y$-axis is $\log_{10}(\bar \xi)$.
For low $ee'$-resonance strength or $\bar \xi \lesssim 10^{-3}$,
the capture probability drops at a drift rate consistent
with the critical value measured 
in section 3 for a single second order resonance
(Equation \ref{eqndb}).
For larger values of $ee'$ resonant term, $\bar \xi$, the critical drift
rate increases, depending upon $\bar \xi$.  In this case the
system fails to capture into the second order $e^2$-resonance and instead
captures into $ee'$-resonance which behaves like a first order resonance.
b)
The Equation (\ref{eqnvarnechi}) was numerically integrated with 
$\bar c_\chi=0$, $\bar\epsilon_\chi = 0$,
and initial momentum $\bar \Gamma (t_0) = 10^{-6}$.
Drift rate and $\bar \chi$ were varied.
The $x$-axis is $\log_{10}|{d \bar b_\chi \over d\tau}|$
and the $y$-axis is $\log_{10}(\bar \chi)$.
Because $\chi = \xi^{-2/3}$ this figure covers large
values of $\xi$, extending past the top of a).
For low $e^2$-resonance strength or  $\bar \xi \lesssim 10^{-1}$,
the capture probability drops at a drift rate consistent
with that predicted in section 3 for a single first order resonance;
(Equation \ref{eqndb}).
For $\bar \chi \sim 0.5$ there is a regime of drift rates with
intermediate capture probability.
\label{fig:xichi}
}
\end{figure*}

\begin{figure*}
\plotone{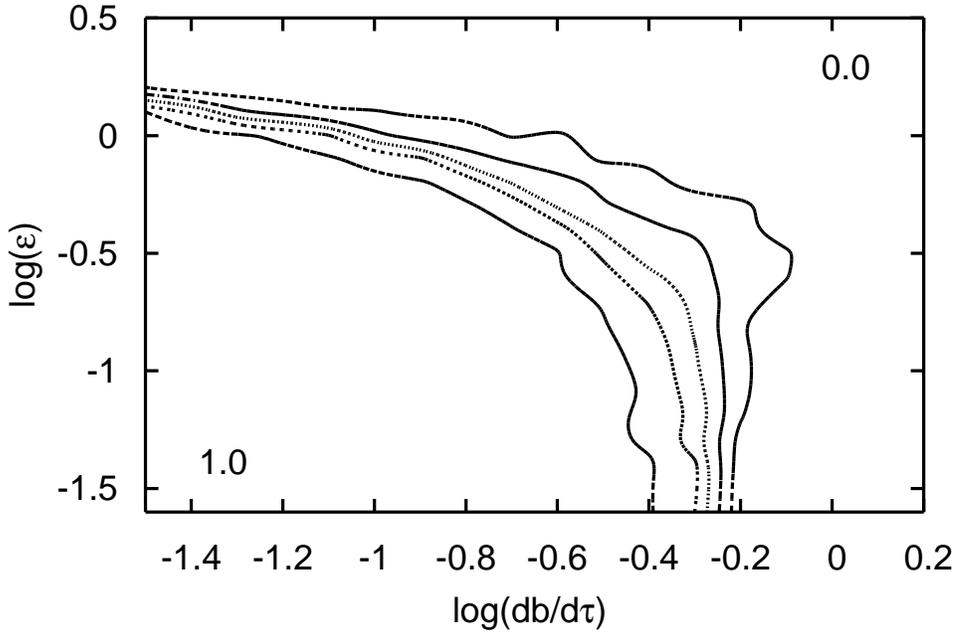}
\figcaption{
The capture probability for a second order resonance ($k=2$)
as a function of drift rate and corotation resonance strength.
This figure is similar to Fig. \ref{fig:a1n}.
The Equation (\ref{eqnvarnexi}) was numerically integrated with 
$\bar c_\xi=0$, $\bar\xi = 0$,
and initial momentum $\bar \Gamma (t_0) = 10^{-6}$.
Drift rate and $\bar \epsilon_\xi$ were varied.
The $x$-axis is $\log_{10}|{d \bar b_\xi \over d\tau}|$
and the $y$-axis is $\log_{10}(\bar \epsilon_\xi)$.
At low values of $\bar \epsilon_\xi$ the capture probability
drops at a value consistent with that predicted in section 3
for a single second order resonance; 
Equation (\ref{eqndb}).
At $\bar \epsilon_\xi \sim 1$ the corotation terms
prevents  capture into the $e^2$-resonance. 
\label{fig:b1n}
}
\end{figure*}

\begin{figure*}
\plottwo{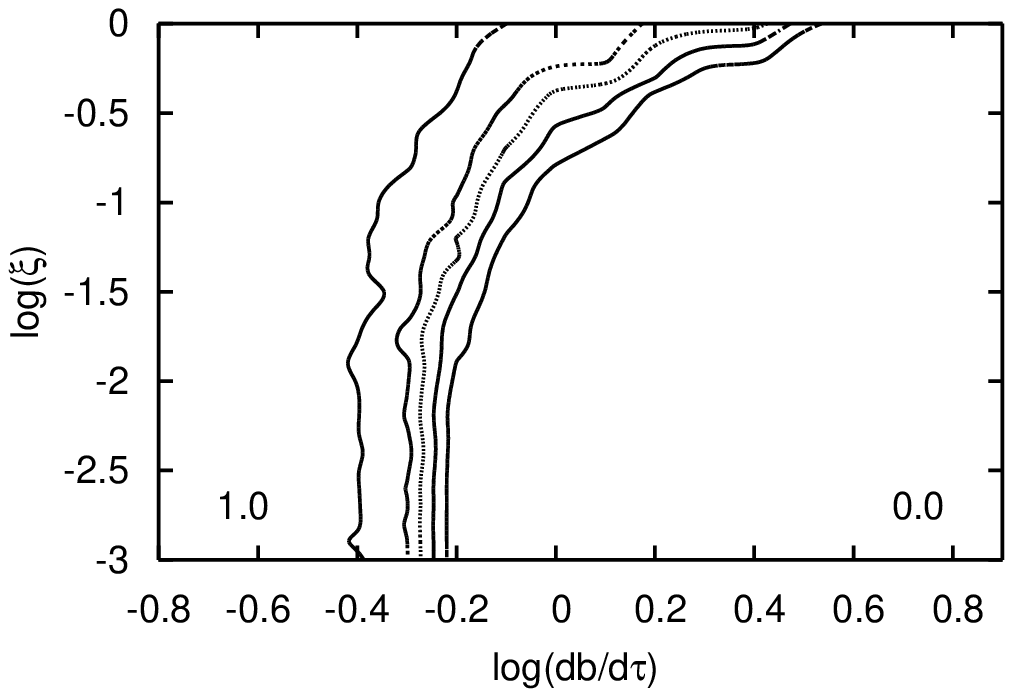}{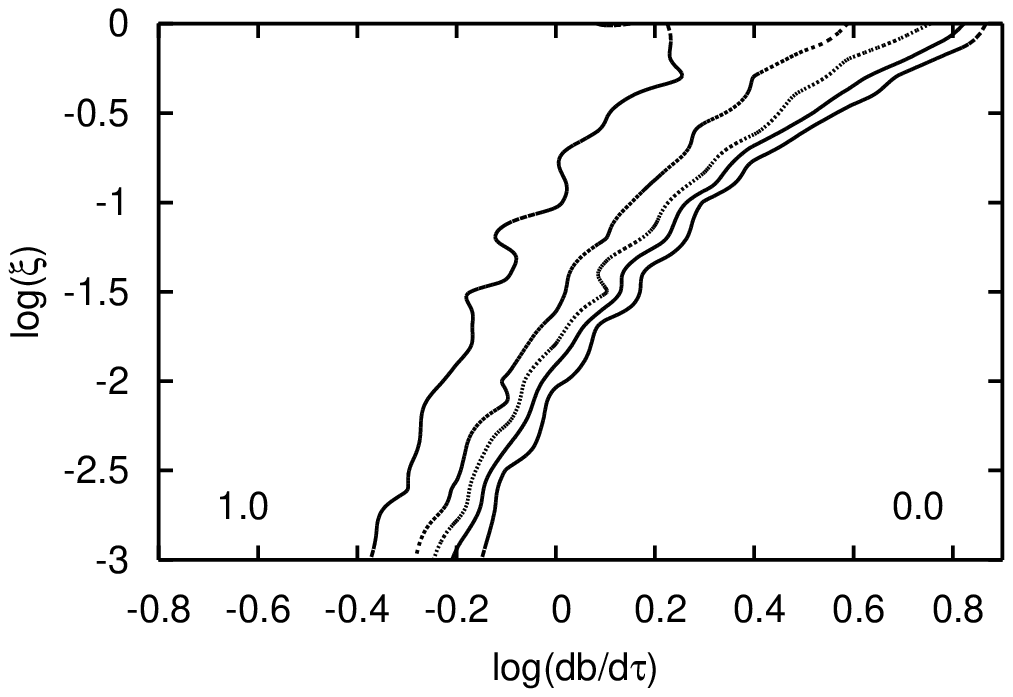}
\figcaption{
Separated second order resonances.
a) This Figure is similar to 
Figure \ref{fig:xichi}a except $\bar c_\xi = -1$.
The $ee'$-resonance is encountered after the $e^2$-resonance.
b) This Figure is similar to 
Figure \ref{fig:xichi}a except $\bar c_\xi =  1$.
The $ee'$-resonance is encountered before the $e^2$-resonance.
\label{fig:xic}
}
\end{figure*}

\begin{figure*}
\plottwo{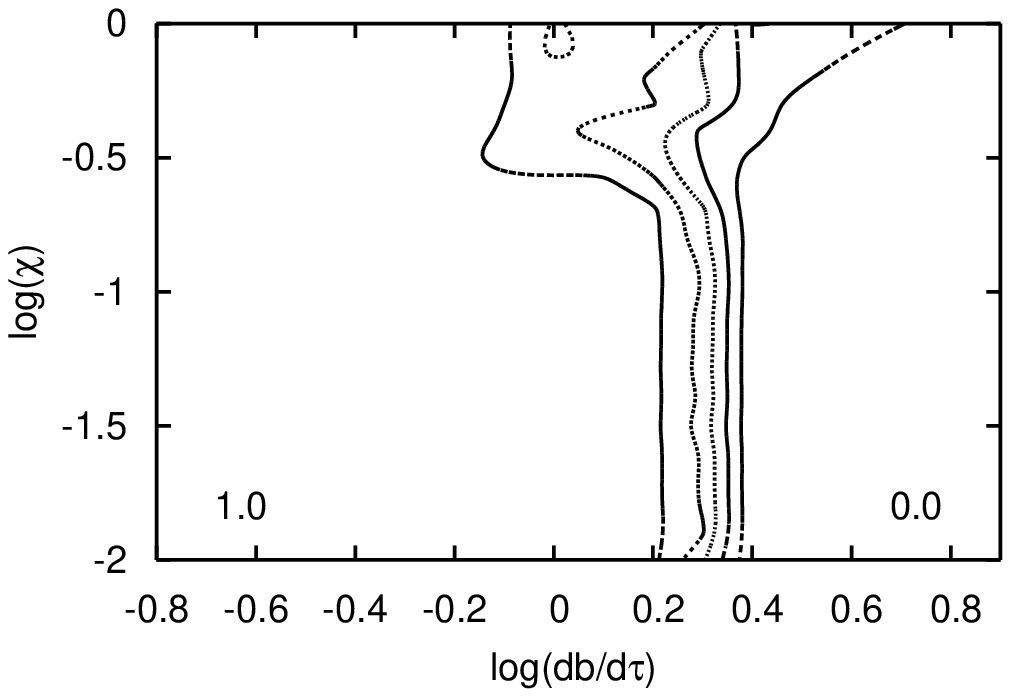}{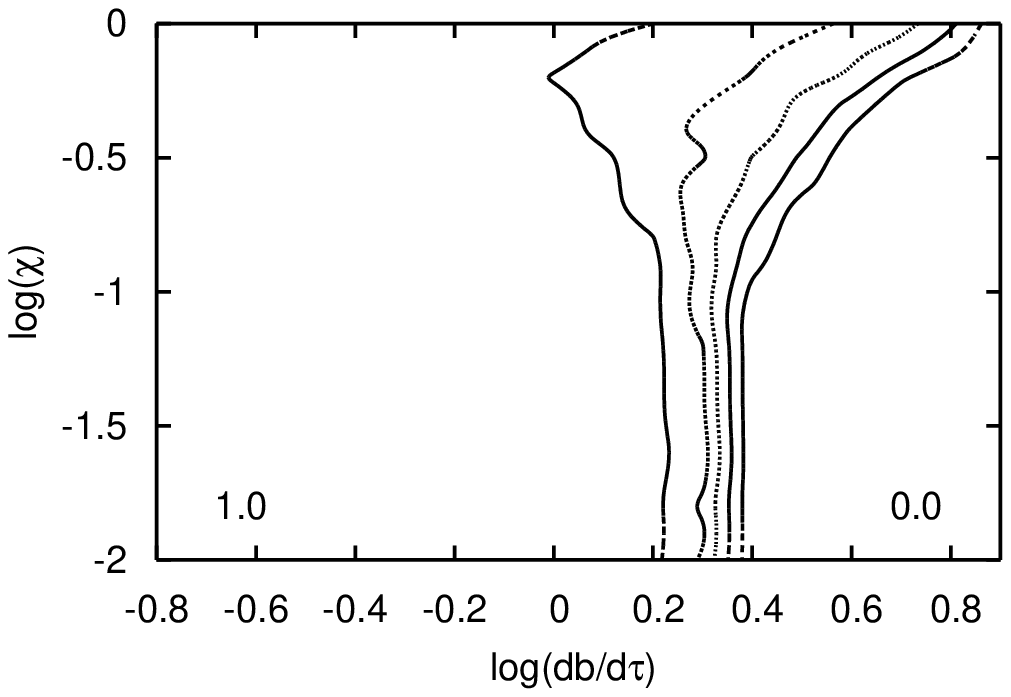}
\figcaption{
Separated second order resonances.
a) This Figure is similar to 
Figure \ref{fig:xichi}b except $\bar c_\chi = -0.5$.
The $ee'$-resonance is encountered before the $e^2$-resonance.
b) This Figure is similar to 
Figure \ref{fig:xichi}b except $\bar c_\chi =  0.5$.
The $ee'$-resonance is encountered after  the $e^2$-resonance.
\label{fig:chic}
}
\end{figure*}

\vfill\eject
\clearpage
\newpage

\begin{deluxetable}{lrrrrrrr}
\tablecolumns{7}
\tablecaption{Coefficients for $k=1$ resonances \label{table:k1}}
\tablehead{
\colhead{} &
\multicolumn{3}{c}{Exterior resonances} &
\multicolumn{3}{c}{Interior resonances} \\
\cline{2-4} \cline{5-7} \\
\colhead{$j:k$} &
\colhead{4:3}   &
\colhead{3:2}   &
\colhead{2:1}   &
\colhead{4:3}   &
\colhead{3:2}   &
\colhead{2:1}   
}
\startdata
$\alpha$            &$ 0.825$ &$ 0.763$   &$ 0.630$   
                    &$ 0.825$ &$ 0.763$   &$ 0.630$   \\ 
$a$                 &$-16.35$ &$-7.86 $   &$-2.38 $   
                    &$-19.81$ &$-10.30$   &$-3.78 $   \\ 
$c/\mu$             &$-4.14 $ &$-2.01 $   &$-0.61 $   
                    &$-5.02 $ &$-2.64 $   &$-0.98 $   \\ 
$\delta_{1,0}/\mu$       
                    &$-3.41 $ &$-2.31 $   &$-0.24 $    
                    &$-4.21 $ &$-3.06 $   &$-1.89 $   \\ 
$\delta_{1,1}/(\mu e_p)$ 
                    &$ 2.35 $ &$ 1.55 $   &$ 0.75 $   
                    &$ 3.07 $ &$ 2.29 $   &$ 0.31 $   \\ 
$\dot n_{p,crit}/\mu^{4/3}$ 
                    & $198.5$ & $48.4$    & $0.54$    
                    & $398.7$ & $126.4$   & $22.7$    \\
$\bar\epsilon/(\mu^{-1/3} e_p)$ 
                    & $1.16$  & $1.00$    & $6.59$    
                    & $1.22$  & $1.21$    & $0.21$    \\
$\bar c/\mu^{1/3}$  & $0.72$  & $0.58$    & $1.18$
                    & $0.71$  & $0.58$    & $0.41$   \\
$e_{lim}/\mu^{1/3}$
                    & $0.98$  & $1.08$    & $0.72$   
                    & $1.08$  & $1.24$    & $1.54$    \\
\enddata
\tablecomments{
Equations (\ref{eqnab}) and (\ref{eqnc}) were used to
calculate the coefficients $a$ and $c$ (for external resonances).  
Equations (\ref{delta1}) and (\ref{delta2to1}) were used
to estimate the perturbation strengths for the 4:3 and 3:2 resonances
and the 2:1 resonance, respectively, for external resonances.
Coefficients for the internal resonances are given in the appendix.
The critical drift rate is calculated from Equation (\ref{eqnnpcrit}).
The coefficients $\bar \epsilon$ and $\bar c$ are 
calculated from Equation (\ref{eqnec}).
We have calculated $c$ using only the secular term from one planet
and assumed that $\dot \varpi_p =0$.
The critical eccentricity ensuring capture in the adiabatic limit
is calculated using Equation (\ref{eqnecrit}).
} 
\end{deluxetable}

\begin{deluxetable}{lrrrr}
\tablecolumns{5}
\tablewidth{0pt}
\tablecaption{Coefficients for $k=2$ resonances \label{table:k2}}
\tablehead{
\colhead{}     &
\multicolumn{2}{c}{Exterior resonances} &
\multicolumn{2}{c}{Interior resonances} \\
\cline{2-3} \cline{4-5} \\
\colhead{$j:k$}     &
\colhead{3:1}   &
\colhead{5:3}   &
\colhead{3:1}   &
\colhead{5:3}   
}
\startdata
%
$\alpha$            &$ 0.481$ &$ 0.711$ 
                    &$ 0.481$ &$ 0.711$   \\
$a$                 &$-3.12 $ &$-18.98$  
                    &$-6.49 $ &$-26.68$   \\
$c/\mu$             &$-0.20 $ &$-1.22 $ 
                    &$-0.41 $ &$-1.72 $   \\
$\delta_{2,0}/\mu$       
                    &$ 0.24 $ &$ 6.82 $   
                    &$ 0.63 $ &$ 3.47 $   \\
$\delta_{2,1}/(\mu e_p)$ 
                    &$-1.25 $ &$-8.00 $  
                    &$-3.76 $ &$-13.33 $  \\
$\delta_{2,2}/(\mu e_p^2)$ 
                    &$0.10$   &$1.04   $ 
                    &$0.36$   &$5.69   $  \\
$\bar\xi/(\mu^{-1/2} e_p)$ 
                    & $18.56$ & $1.95 $   
                    & $19.11$ & $10.61 $  \\
$\bar \epsilon_\xi /(\mu^{-1} e_p^2)$    
                    & $5.59 $ & $0.43 $ 
                    & $5.93 $ & $12.56$   \\
$\bar c_\xi /\mu^{0}$   
                    & $0.81$  & $0.18 $ 
                    & $0.65$  & $0.50 $   \\
$\dot n_{p,crit,\xi}/\mu^{2}$ 
                    & $0.029$ & $69.86$    
                    & $0.60$  & $30.19$   \\
$e_{lim,\xi}/\mu^{1/2}$
                    & $0.40$  & $0.95$    
                    & $0.19$  & $0.20$    \\
$\bar \epsilon_\chi /(\mu^{-1/3} e_p^{2/3})$    
                    & $0.114$ & $0.174$ 
                    & $0.116$ & $0.538$   \\
$\bar c_\chi/(\mu^{1/3} e_p^{-2/3})$ 
                    & $0.12$  & $0.11 $
                    & $0.09$  & $0.10 $   \\
$\dot n_{p,crit,\chi}/(\mu^{4/3} e_p^{4/3})$ 
                    & $5.77$  & $683$     
                    & $40.66$ & $1694$    \\
$e_{lim,\chi}/(\mu^{1/3}e_p^{1/3})$
                    & $1.06$  & $1.19$    
                    & $1.73$  & $1.50$    \\
\enddata
\tablecomments{
Equations (\ref{eqnab}) and (\ref{eqnc}) were used to
calculate $a$ and $c$.  
The critical drift rates 
$\dot n_{p,crit,\xi}$ and $\dot n_{p,crit,\chi}$ are those
calculated  for low $\bar \xi$ and $\bar \chi$ respectively.
$\dot n_{p,crit,\xi}$ is given in the limit of low initial 
particle eccentricity.
For particle initial eccentricity near the limit ensuring capture
in the adiabatic limit, the values given here for $n_{p,crit,\xi}$ should
be multiplied by 10.
The coefficients $\bar \epsilon_\xi$, and $\bar c_\xi$ are 
calculated from Equation (\ref{eqnxi}).
The coefficients $\bar \epsilon_\chi$, and $\bar c_\chi$ are 
calculated from Equation (\ref{eqnchi}).
Expressions from the appendix were used to calculate the $\delta$ 
coefficients and critical eccentricities. 
We have calculated $c$ using only the secular term from one planet
and assumed that $\dot \varpi_p =0$.
} 
\end{deluxetable}

\end{document}